%% file: main.tex
\newif\ifshortversion
\newif\ifenversion

\enversiontrue % Comment this to compile the portuguese version

\ifshortversion
	\documentclass[runningheads]{llncs}
	\renewcommand{\keywordname}{{\bfseries Palavras-chave:}}
\else
	\documentclass[a4paper,UKenglish,cleveref,autoref]{lipics-v2019}
\fi

\ifshortversion
\ifenversion
\usepackage[english]{babel}
\else
\usepackage[portuguese]{babel}
\fi
\usepackage{graphicx}
\else
\fi
\usepackage[utf8]{inputenc}
\usepackage{listingsutf8}
\usepackage{verbatim}
\usepackage{why3lang}
\usepackage{amsmath}
\usepackage{hyperref}
\usepackage[htt]{hyphenat}

\ifenversion
\newcommand{\alglong}[0]{conversion algorithm to CNF}
\else
\newcommand{\alglong}[0]{algoritmo de conversão para FNC}
\fi

\newcommand{\implfree}[0]{{\texttt{Impl\_Free}}}
\newcommand{\nnfc}[0]{{\texttt{NNFC}}}
\newcommand{\cnfc}[0]{{\texttt{CNFC}}}

\newcommand{\mypg}[1]{\textit{\textbf{#1}.}}
\newcommand{\myparagraph}[1]{\medskip \noindent \textit{#1}.}
\newcommand{\myparagraphbf}[1]{\medskip \noindent \textit{\textbf{#1}.}}

\ifenversion
\title{Animated Logic: Correct Functional Conversion to Conjunctive Normal Form}
\else
\title{Lógica animada: conversão funcional correta para Forma Normal Conjuntiva\thanks{Este trabalho  é financiado pela Fundação Tezos através do
projeto FACTOR e, por fundos nacionais, através da FCT – Fundação para a Ciência e a Tecnologia, I.P., no âmbito do NOVA LINCS através do projeto UID/CEC/04516/2019.}\thanks{Aceite e apresentado no INForum 2019 - Simpósio de Informática -- http://inforum.org.pt/}}
\fi

\ifshortversion
    \author{Pedro Barroso,
    Mário Pereira e
    António Ravara}
    
    \authorrunning{}
    \institute{NOVA LINCS \& DI-FCT -- Universidade Nova de Lisboa, Portugal}
\else
    \author{Pedro Barroso, Mário Pereira and António Ravara}{NOVA LINCS \& Departamento de Informática, Faculdade de Ciências e Tecnologias da Universidade Nova de Lisboa, Portugal}{}{}{}%TODO mandatory, please use full name; only 1 author per \author macro; first two parameters are mandatory, other parameters can be empty. Please provide at least the name of the affiliation and the country. The full address is optional
    
    \authorrunning{P. Barroso, M. Pereira and A. Ravara}%TODO mandatory. First: Use abbreviated first/middle names. Second (only in severe cases): Use first author plus 'et al.'
    
    \Copyright{Pedro Barroso, Mário Pereira and António Ravara}%TODO mandatory, please use full first names. LIPIcs license is "CC-BY";  http://creativecommons.org/licenses/by/3.0/
    
    \ccsdesc[100]{Theory of computation~Logic and verification}
    \keywords{Computational logic, conjunctive normal form, functional programming, proof of programs, Why3}%TODO mandatory; please add comma-separated list of keywords
    
    \category{}%optional, e.g. invited paper
    
    \relatedversion{}%optional, e.g. full version hosted on arXiv, HAL, or other respository/website
    \supplement{}%optional, e.g. related research data, source code, ... hosted on a repository like zenodo, figshare, GitHub, ...
    
    \funding{Tezos Foundation}%optional, to capture a funding statement, which applies to all authors. Please enter author specific funding statements as fifth argument of the \author macro.
    
    \acknowledgements{This work is funded by Tezos Foundation through the FACTOR project.}
    
    \nolinenumbers %uncomment to disable line numbering
    
    \hideLIPIcs  %uncomment to remove references to LIPIcs series (logo, DOI, ...), e.g. when preparing a pre-final version to be uploaded to arXiv or another public repository
    
    \EventEditors{John Q. Open and Joan R. Access}
    \EventNoEds{2}
    \EventLongTitle{42nd Conference on Very Important Topics (CVIT 2016)}
    \EventShortTitle{CVIT 2016}
    \EventAcronym{CVIT}
    \EventYear{2016}
    \EventDate{December 24--27, 2016}
    \EventLocation{Little Whinging, United Kingdom}
    \EventLogo{}
    \SeriesVolume{42}
    \ArticleNo{23}
\fi

\begin{document}
\maketitle              % typeset the header of the contribution
%

%Table of contents; only for long version
\ifshortversion
\else
%\tableofcontents
\fi

\ifenversion
\input{English/abstractEN.tex}

\input{English/introductionEN.tex}
\input{English/descriptionEN.tex}
\input{English/criteriaEN.tex}
\input{English/directstyleEN.tex}
\input{English/cpsEN.tex}
\input{English/desfuncionalizationEN.tex}
\input{English/conclusionEN.tex}
\else
\input{abstract}
\input{introducao}
\input{descricao}
\input{criterios}
\input{estilodireto}
\input{cps}
\input{desfuncionalizacao}
\input{conclusao}
\fi

\bibliographystyle{splncs04}
\bibliography{main}

\appendix
\ifenversion
\input{English/appendixEN.tex}
\else
\include{appendice}
\fi
\end{document}

%% file: English/abstractEN.tex
\begin{abstract}
%In this article, w
We present an approach to %the
obtain formally verified implementations of classical Computational Logic algorithms. We choose %as a testing tool
the Why3 platform because it %that 
allows to implement %near
functions in a style very close to the mathematical definitions, as well as it allows a high degree of automation in the verification process.

As proof of concept, we present a mathematical definition of the algorithm to convert propositional formulae to conjunctive normal form, implementations in WhyML (the Why3 language, very similar to OCaml), and proofs of correctness of the implementations. We apply our proposal on two variants of this algorithm: one in direct-style and another with an explicit stack structure. Being both first-order versions, Why3 processes the proofs naturally.
\end{abstract}

%% file: English/introductionEN.tex
\section{Introduction}
\label{sec:motivation}

\myparagraphbf{Motivation} Foundational courses in Computer Science, like Computational Logic, aim at presenting key basilar subjects to the education of undergraduate students. To strength the relation of the topics covered to sound programming practices, it is relevant to link the mathematical content to clear and executable implementations, provably correct to stress the importance of sound practices.

Herein we present work developed in the context of the \texttt{FACTOR} \cite{factor} project, which aims to promote the use of OCaml \cite{leroy2014ocaml} and correct code development practices in the Portuguese-speaking academic community. Specifically, the objectives of the project are the functional implementation of classical Computational Logic algorithms and Formal Languages, the accomplishment of correctness proofs and the step-by-step execution to help understanding the algorithm.

The algorithm for converting propositional formulae to Conjunctive Normal Form (CNF)\footnote{A formula is in CNF if it is a conjunction of clauses, where a clause is a disjunction of literals and a literal is a propositional symbol or its negation.} is often presented formally, with rigorous mathematical definitions that are sometimes difficult to read \cite{inttologicenderton, logichamilton, mathlogicmendelson}, or informally, intended for Computer Science but with textual definitions in non-executable pseudo-code \cite{mathlogicbenari, huth2004logic}. The implementation of algorithms of this nature is a fundamental piece for learning and understanding them. Languages such OCaml allow implementations very close to the mathematical definitions, helping the study because they are executable. Also, the correctness proof of a functional implementation is simpler than for the imperative one.

\myparagraphbf{Contributions} As proof of concept, we implement and prove correct the referred algorithm in Why3 \cite{filliatre2013why3}, a platform for deductive program verification. Why3 provides a first-order language with polymorphic types, pattern matching and inductive predicates, called WhyML. Also offers a certified OCaml code extraction mechanism and support for third-party provers.

To support the step-by-step execution of the algorithm, an important feature to help students understanding the definitions, we also implemented a version in Continuation-Passing Style (CPS) \cite{sabry1993reasoning} and via defunctionalization got an evaluator, a version close to a first-order abstract machine~\cite{danvy2003functionalAM}. Due to the limited support of Why3 to the higher order, it was not possible to close the correctness proof for the CPS version. This limitation however is not present in the defunctioned implementation that has an explicit stack structure, but in first-order. This implementation resulted from a mechanical transformation from the CPS version. This version has been naturally proven correct by Why3.

%\myparagraphbf{Contributions} T
In short, this article presents pedagogical material to support the teaching of classical Computational Logic algorithms. We developed two implementations, formally verified in Why3, from a presentation as a recursive function of \alglong: the first in direct style and the second with an explicit stack structure. Both were proved sound with small effort, basically following from the assertions one naturally associates with the code to prove it correct.

%Since from these implementations it is possible to extract OCaml code.

%% file: English/descriptionEN.tex
\newcommand\myeq{\mathrel{\stackrel{\makebox[0pt]{\mbox{\normalfont\tiny abv}}}{=}}}

\section{Functional presentation of the algorithm}
\label{sec:algdesc}

\mypg{Description} Let us call T to the algorithm that converts any propositional logic formula to CNF. A propositional formula $\phi$ is an element of the set $G_p$, defined as follows:
$$
G_p \triangleq \ \phi ::= T \ | \ p \ | \ \neg \phi \ | \ \phi \wedge \phi \ | \ \phi \vee \phi \ | \ \phi \rightarrow \phi \ 
$$
The function T produces formula without the implication connective, so we define also the set $H_p$ as a subset of $G_p$ without implications. We thus have T: $G_p \rightarrow H_p$, where:

%MISSING GP AND HP Description

%Chamamos T ao algoritmo que converte qualquer fórmula de Lógica Proposicional para FNC. Dada uma fórmula proposicional~$\phi$, o conjunto $G_p$ de fórmulas proposicionais com os conetivos habituais e o o conjunto $H_p \subseteq G_p$ que não contem o conectivo de implicação, temos T: $G_p \rightarrow H_p$, onde

%, fazendo primeiro a eliminação das implicações (\texttt{Impl\_Free}), posterior eliminação de duplas negações e transformação para a forma normal de negação \footnote{Uma fórmula $\phi \in H_p$ diz-se que está na forma normal da negação, se só as suas sub-fórmulas que são fórmulas atómicas estão negadas.} (\texttt{NNFC}) e finalizando com a transformação para a FNC (\texttt{CNFC}). Chamando-lhe T  para uma fórmula $\phi$, onde T: $G_p \rightarrow H_p$, temos: %Adicionar o nome das funcionais
\begin{center}
T($\phi$) = CNFC(NNFC(Impl\_Free($\phi$))) 
\end{center}
%$G_p \triangleq \ \phi ::= T \ | \ p \ | \ \neg \phi \ | \ \phi \wedge \phi \ | \ \phi \vee \phi \ | \ \phi \rightarrow \phi \ $

%$H_p \subseteq G_p$ is the set of formulae without implications

\noindent The algorithm composes three functions: 

\begin{itemize}
    \item The {\implfree} function responsible for eliminating the implications;
    \item The {\nnfc} function responsible for converting to Negation Normal Form (NNF)\footnote{A formula is in NNF if the negation operator is only applied to sub-formulae that are literals.}
    \item The {\cnfc} function responsible for converting from NNF to CNF.
\end{itemize}

\myparagraphbf{Implementation}
\label{subsec:implalg}
To represent the set $G_p$, we define the type \texttt{formula} that declares variables (\texttt{FVar}) and constants (\texttt{FConst}), and constructs formulas with conjunctions (\texttt{FAnd}), disjunctions (\texttt{FOr}), implications (\texttt{FImpl}) and negations (\texttt{FNeg}):

\begin{why3}
type formula =
  | FVar ident
  | FConst bool
  | FAnd formula formula
  | FOr formula formula
  | FImpl formula formula
  | FNeg formula
\end{why3}
%Estrategicamente escolhemos internalizar a ausência de implicações no tipo \texttt{formula\_wi}.
To represent the $H_p$ set we define the type \texttt{formula\_wi}, similar to the previous but without the implication constructor.

%A ausência de conetivos de implicação é assegurado pelo tipo de retorno da função (\texttt{formula\_wi});

%%%%%%%%%%%% MAYBE PRESENT THE SET

\medskip
\noindent We present now the implementation of the three functions.

The function \texttt{Impl\_Free} removes all the implications. It is recursively defined in the cases of the type \texttt{formula} and homomorphic, except in the implication case where it takes advantage of the Propositional Logic Law: 

$$
\mathtt{A} \rightarrow \mathtt{B} \equiv \neg \mathtt{A} \vee \mathtt{B}
$$
It converts the constructions of the type \texttt{formula} for those of the type \texttt{formula\_wi} and does recursive calls over the arguments:

\begin{why3}
let rec impl_free (phi: formula) : formula_wi
= match phi with
  | FNeg phi1 -> FNeg_wi (impl_free phi1)
  | FOr phi1 phi2 -> FOr_wi (impl_free phi1) (impl_free phi2)
  | FAnd phi1 phi2 -> FAnd_wi (impl_free phi1) (impl_free phi2)
  | FImpl phi1 phi2 -> FOr_wi (FNeg_wi (impl_free phi1)) (impl_free phi2)
  | FConst phi -> FConst_wi phi
  | FVar phi -> FVar_wi phi
end
\end{why3}

The functions {\nnfc} converts formulas to NNF. It is recursively defined over a combination of constructors:  applying the Propositional Logic Law \texttt{$\neg \neg$A $\equiv$ A} the double negations are eliminated and using the De Morgan Laws, negations of conjunctions become disjunction of negations and negations of disjunctions become conjunction of negations. The code of the function is as follows:
    
\begin{why3}
let rec nnfc (phi: formula_wi) : formula_wi
= match phi with
  | FNeg_wi (FNeg_wi phi1) -> nnfc phi1
  | FNeg_wi (FAnd_wi phi1 phi2) -> FOr_wi (nnfc (FNeg_wi phi1)) 
    (nnfc (FNeg_wi phi2))
  | FNeg_wi (FOr_wi phi1 phi2) -> FAnd_wi (nnfc (FNeg_wi phi1)) 
    (nnfc (FNeg_wi phi2))
  | FOr_wi phi1 phi2 -> FOr_wi (nnfc phi1) (nnfc phi2)
  | FAnd_wi phi1 phi2 -> FAnd_wi (nnfc phi1) (nnfc phi2)
  | phi -> phi
end
\end{why3}

The {\cnfc} function converts formulas from NNF to CNF. It is straightforwardly defined except in the disjunction case, where it distributes the disjunction by the conjunction calling the auxiliary function \texttt{distr}.

\begin{why3}
let rec cnfc (phi: formula_wi) : formula_wi
= match phi with
  | FOr_wi phi1 phi2 -> distr (cnfc phi1) (cnfc phi2)
  | FAnd_wi phi1 phi2 -> FAnd_wi (cnfc phi1) (cnfc phi2)
  | phi -> phi
end
\end{why3}

The \texttt{distr} function uses the Propositional Logic Law
$$\mathtt{A} \vee (\mathtt{B} \wedge \mathtt{C}) \equiv (\mathtt{A} \vee \mathtt{B}) \wedge (\mathtt{A} \vee \mathtt{C}),$$
the code is the following:
\begin{why3} 
let rec distr (phi1 phi2: formula_wi) : formula_wi
= match phi1, phi2 with
  | FAnd_wi phi11 phi12, phi2 -> FAnd_wi (distr phi11 phi2) 
    (distr phi12 phi2)
  | phi1, FAnd_wi phi21 phi22 -> FAnd_wi (distr phi1 phi21) 
    (distr phi1 phi22)
  | phi1,phi2 -> FOr_wi phi1 phi2
end
\end{why3}

Lastly, the code of the function (\texttt{T}) %that 
composes all of these functions:
\begin{why3}
let t (phi: formula) : formula_wi 
= cnfc(nnfc(impl_free phi))
\end{why3}
From this WhyML implementation it is possible to extract %the
OCaml code (Appendix \ref{appendix:codigoocaml}).
%In both implementations, the similarity with 
Both implementations are close to 
the mathematical definitions %emphasizes,
thus demonstrating that functional languages like OCaml are suitable languages for the presentation of these algorithms, providing executable definitions without sacrificing rigor or clarity.

%% file: English/criteriaEN.tex
\section{How to obtain the correctness}%Correção da Implementação}
\label{sec:corrcrite}

%The 
Since the \texttt{T} algorithm %, as previously shown, 
is a composition of three functions, %. The
the correctness of the algorithm is the result of the correctness criteria of each of these three functions.

\myparagraphbf{Criteria}
\label{sec:corrcrite} %reformular e introduzir pós e pré-condições, perguntar.
For all functions the basic correctness criterion is that the input and output formula must be logically equivalent. In addition it is required that:

\begin{itemize}
    \item[--] \textbf{Impl\_Free:}
    \begin{itemize}
        \item The result should not contain implications.% connectives.
    \end{itemize}
    \item[--] \textbf{NNFC:}
    \begin{itemize}
        \item The input and output formula should not contain implications.% connectives.
    	\item The result must be in %the negation normal form.
    	Negation Normal Form.
    \end{itemize}
    \item[--] \textbf{CNFC:}
    \begin{itemize}
    	\item The input and output formula should not contain implications.% connectives.
    	\item The input and output formula must be in %the negation normal form.
    	Negation Normal Form.
    	\item The result must be in %the conjunctive
    	Conjunctive Normal Form.
    \end{itemize}
\end{itemize}

\noindent %Note that the
The correctness criteria of a function %are
needs to be propagated to the following ones, to ensure that each does not violate the conditions already %assured
established by the one previously executed. %functions.

\myparagraphbf{Semantics of formulae}
Since the basic criterion of correctness is the logical equivalence of formulae, we need a function to assign a semantic to them:

\begin{why3}
type valuation = ident -> bool

function eval (v: valuation) (f: formula) : bool
= match f with
  | FVar x      -> v x
  | FConst b    -> b
  | FAnd f1 f2  -> eval v f1 /\ eval v f2
  | FOr f1 f2   -> eval v f1 || eval v f2
  | FImpl f1 f2 -> (eval v f1 -> eval v f2)
  | FNeg f      -> not (eval v f)
end
\end{why3}
This function takes an argument of type \texttt{valuation} assigning a value of type \texttt{bool}\footnote{bool is the Boolean type of WhyML.} to each variable of the formula, receives the formula to %be evaluated 
evaluate and returns a value of type \texttt{bool}.
For the base constructors \texttt{FVar} and \texttt{FConst}, the Boolean value of the variable and the value of the constant, respectively, are returned. For the remaining constructor cases, the associated formulae are recursively evaluated and the result translated into the corresponding WhyML Boolean operation. The valuation function for the type of formulae \texttt{formula\_wi} is similar.

%% file: English/directstyleEN.tex
\section{Proof of correctness}

The proof of correctness consists in demonstrating that each function respects the correctness criteria defined in the previous section. We show herein the WhyML code accepted by Why3 as correct.

\myparagraphbf{Correctness of Impl\_Free}
The absence of implication connectives is ensured by the return type of the function (\texttt{formula\_wi}); the equivalence of the formulae is ensured using the formula valuation functions and we use the input formula as a measure to ensure termination.

\begin{why3}
let rec function impl_free (phi: formula) : formula_wi
  ensures { forall v. eval v phi = eval_wi v result }
  variant { phi }
= ...
\end{why3}

\myparagraphbf{Correctness of NNFC} %talvez mudar FNN para forma normal de negação
The absence of implication connectives in the input and output formulae is ensured by the type \texttt{formula\_wi}. Additionally, to prove that the result is in the NNF, we introduce %the well-formulated
a well-formedness predicate \texttt{wf\_negations\_of\_literals}, which states that the sub-formulae of the constructor \texttt{FNeg\_wi} cannot contain the constructors \texttt{FOr\_wi}, \texttt{FAnd\_wi}, or \texttt{FNeg\_wi}:

\begin{why3}
predicate wf_negations_of_literals (f: formula_wi)
= match f with
  | FNeg_wi f -> (forall f1 f2. f <> FOr_wi f1 f2 /\ f <> FAnd_wi f1 f2 /\ f <> FNeg_wi f1) /\ wf_negations_of_literals f
  | FOr_wi f1 f2 | FAnd_wi f1 f2 -> wf_negations_of_literals f1 /\ wf_negations_of_literals f2
  | FVar_wi _ -> true
  | FConst_wi _ -> true
end
\end{why3}
In %this prove
the proof of correctness it is not possible to use the formula itself as a measure of termination, since in the case of the distribution of negation by conjunction or disjunction, constructors are added to the head constructors, making the structural inductive criterion %impossible
not applicable. %We then created 
Hence, we define a function that counts the number of constructors of each formula and use it as termination measure:

\begin{why3}
function size (phi: formula_wi) : int 
= match phi with
  | FVar_wi _ | FConst_wi _ -> 1
  | FNeg_wi phi -> 1 + size phi
  | FAnd_wi phi1 phi2 | FOr_wi phi1 phi2 -> 1 + size phi1 + size phi2
end
\end{why3}
%However, it is necessary t
To ensure the number of constructors can never be negative, %. For this, 
we use an auxiliary lemma:

\begin{why3}
let rec lemma size_nonneg (phi: formula_wi)
  variant { phi }
  ensures { size phi >= 0 }
= match phi with
  | FVar_wi _ | FConst_wi _ -> ()
  | FNeg_wi phi -> size_nonneg phi
  | FAnd_wi phi1 phi2 | FOr_wi phi1 phi2 ->
        size_nonneg phi1; size_nonneg phi2
end
\end{why3}
So, now with the well-formedness predicate and termination measure defined, we can close the proof of correctness of the NNFC function: %The code submitted to the proof is the following:

\begin{why3}
let rec nnfc (phi: formula_wi) : formula_wi
  ensures { forall v. eval_wi v phi = eval_wi v result }
  ensures { wf_negations_of_literals result }
  variant { size phi }
= ...
\end{why3}

\myparagraphbf{Correctness of CNFC} To ensure that a given formula is in CNF we introduce the well-formedness predicates \texttt{wf\_conjunctions\_of\_disjunctions} and \texttt{wf\_disjunctions}. These guarantee that after a disjunction there are no conjunctions:

\begin{why3}
predicate wf_conjunctions_of_disjunctions (f: formula_wi)
= match f with
  | FAnd_wi f1 f2 -> wf_conjunctions_of_disjunctions f1 /\ wf_conjunctions_of_disjunctions f2
  | FOr_wi f1 f2 -> wf_disjunctions f1 /\ wf_disjunctions f2
  | FConst_wi _ -> true
  | FVar_wi _ -> true
  | FNeg_wi f1 -> wf_conjunctions_of_disjunctions f1
end

predicate wf_disjunctions (f: formula_wi)
= match f with
  | FAnd_wi _ _ -> false
  | FOr_wi f1 f2 -> wf_disjunctions f1 /\ wf_disjunctions f2
  | FConst_wi _ -> true
  | FVar_wi _ -> true
  | FNeg_wi f1 -> wf_disjunctions f1
end
\end{why3}
Lastly, we add the predicates \texttt{wf\_conjunctions\_of\_disjunctions} and \texttt{wf\_negations\_of\_literals} to the post-conditions, to ensure that the result is in NNF and CNF, respectively; to ensure that the input formula is in NNF, we also add the predicate \texttt{wf\_negations\_of\_literals} to the pre-conditions. The code is as follows:

%Adicionalmente, é preciso provar que o resultado respeita a forma normal conjuntiva e assegurar que a fórmula de entrada respeita a forma normal de negação. Para o primeiro caso foi criado o predicado de boa formação \texttt{wf\_conjunctions\_of\_disjunctions} que assegura que depois de uma disjunção não se pode encontrar nenhuma conjunção, o código está no Apêndice \ref{appendix:cnf}; para o segundo basta adicionar o predicado de boa formação \texttt{wf\_negations\_of\_literals} às pré-condições da função.
\begin{why3}
let rec cnfc (phi: formula_wi)
  requires { wf_negations_of_literals phi }
  ensures  { forall v. eval_wi v phi = eval_wi v result }
  ensures  { wf_negations_of_literals result }
  ensures  { wf_conjunctions_of_disjunctions result } 
  variant  { phi }
= ...
\end{why3}
Since %\texttt{distr} is an auxiliary function of 
the \texttt{CNFC} function uses the auxiliary function \texttt{distr}, we also need to prove its correctness. %In this function i
We need to ensure the same criteria of the \texttt{CNFC} function, but because it is
the distribution of the disjunctions by the conjunctions, we must additionally ensure that the
input formulae are in the CNF, which is obtained by adding the predicates
\texttt{wf\_conjunctions\_of\_disjunctions} and \texttt{wf\_negations\_of\_literals} to the
pre-conditions:

\begin{why3}
let rec distr (phi1 phi2: formula_wi)
  requires { wf_negations_of_literals phi1 }
  requires { wf_negations_of_literals phi2 }
  requires { wf_conjunctions_of_disjunctions phi1 } 
  requires { wf_conjunctions_of_disjunctions phi2 }
  ensures  { forall v. eval_wi v (FOr_wi phi1 phi2) = eval_wi v result }
  ensures  { wf_negations_of_literals result }
  ensures  { wf_conjunctions_of_disjunctions result }
  variant  { size phi1 + size phi2 }
= ...
\end{why3}
However in the \texttt{distr} function, it is not possible to prove that a disjunction of two
formulae in CNF is effectively a formula in CNF, because we must ensure that in a disjunction of
two formulae in CNF, the formulae do not contain the constructor \texttt{FAnd\_wi}. To
accomplish this, we %reinforce the proof with
use an auxiliary lemma:

\begin{why3}
lemma aux: forall x. wf_conjunctions_of_disjunctions x ->
  wf_negations_of_literals x -> not (exists f1 f2. x = FAnd_wi f1 f2) ->
  wf_disjunctions x
\end{why3}

\myparagraphbf{Correctness of T}
\label{subsec:provat}
With the proofs of correctness of each of the three functions performed, we can now obtain the proof of correctness of the function \texttt{T}. This guarantees the criteria ensured by the three functions: it does not contain implications due to the return type (\texttt{formula\_wi}); is in NNF (line 2); is in CNF (line 3); the result is equivalent to the input formula (line 4):

%comentário sobre cada ensures
\begin{why3}
let t (phi: formula) : formula_wi
  ensures { wf_negations_of_literals result }
  ensures { wf_conjunctions_of_disjunctions result }
  ensures { forall v. eval v phi = eval_wi v result  }
= cnfc (nnfc (impl_free phi))
\end{why3}
The proof of this "direct style" implementation %and
specification - close to classical mathematical definitions - is immediate in Why3, making this exercise a successful proof of concept: classical logical algorithms presented as functions to undergraduates can have a very close functional implementation that is easy to prove correct with a high degree of automation.

%% file: English/cpsEN.tex
\section{Continuation-Passing Style}

\textit{Continuation-Passing Style} (CPS) is a programming style where the control is passed explicitly in the form of a continuation, thus avoiding the overflow of the stack if the underlying compiler optimizes recursive terminal calls. With an explicit stack structure in the code, it is possible in the future to introduce a mechanism that allows step-by-step execution of the functions.

\myparagraphbf{Process transformation into CPS}
%Manter os argumentos, reescrever
%Dado uma função do t' -> t vamos construir uma nova função de t para um novo
The transformation is performed mechanically according to the following steps:

\begin{itemize}
    \item Given a function of type \texttt{t' $\rightarrow$ t}, we add an argument %will be added 
    which %will 
    represents the continuation (a function of type \texttt{t $\rightarrow$ 'a}) and change the return type of the function to \texttt{'a}.
    \item For the base cases instead of returning the desired values, we apply these values to the continuation function.
    \item For the remaining cases, we start by making a recursive call and construct the continuations with the rest of the computation.
    \item We add a \texttt{main} function that calls the function in CPS with the identity function as a continuation.%Creation
\end{itemize}

%Applying
We apply now this process to the functions presented in the previous section. With respect to the \texttt{Impl\_Free} function:
\begin{enumerate}
    \item We add an argument to the function of type \texttt{formula\_wi $\rightarrow$ 'a} and change the return type to \texttt{'a}:
    \begin{why3}
  let rec impl_free_cps (phi: formula) (k: formula_wi -> 'a ) : 'a
    \end{why3}
    \item For the base cases, we apply the desired values to the continuation function.
    \begin{why3}
    | FConst phi -> k (FConst_wi phi)
    | FVar phi -> k (FVar_wi phi)
    \end{why3}
    \item For the remaining cases, we start with a recursive call and define the continuations with the rest of the computation: \begin{why3}| FNeg phi1 -> impl_free_cps phi1 (fun con -> k (FNeg_wi con))
    | FOr phi1 phi2 -> impl_free_cps phi1 (fun con -> impl_free_cps phi2 (fun con1 -> k (FOr_wi con con1)))
    | FAnd phi1 phi2 -> impl_free_cps phi1 (fun con -> impl_free_cps phi2 (fun con1 -> k (FAnd_wi con con1)))
    | FImpl phi1 phi2 -> impl_free_cps phi1 (fun con -> impl_free_cps phi2 (fun con1 -> k (FOr_wi (FNeg_wi con) con1)))\end{why3}
    \item Finally, we create the %Creation of
    \texttt{main} function that calls the function in CPS with the identity function as a continuation:
    \begin{why3}
  let impl_free_main (phi: formula) : formula_wi
  = impl_free_cps phi (fun x -> x)\end{why3}
\end{enumerate}

We obtain %the implementation of 
the CPS version of the remaining functions in a similar way to this process (the complete code is in Appendix \ref{appendix:cps}).

\myparagraphbf{Correctness criteria% specification
}
One interesting aspect of the proof of correctness of the functions in CPS is the use of the corresponding function in direct style, since these are pure and total functions, as %own
specification, \emph{i.e.}, we simply assure that the result is equal to the result of the functions in direct style, applied to the continuation.

For the {\implfree} function in CPS, it is enough to ensure that the result is equal to the result of the direct-style {\implfree} function applied to the continuation:

\begin{why3}
let rec impl_free_cps (phi: formula) (k: formula_wi -> 'a ) : 'a
  variant { phi }
  ensures { result = k(impl_free phi) }
= ...
\end{why3}
The specification of the function in direct style is then also applied to the function \texttt{main}, responsible for calling the CPS functions with the identity function as a continuation:
\begin{why3}
let impl_free_main (phi: formula) : formula_wi
  ensures { forall v. eval v phi = eval_wi v result }
= ...
\end{why3}
The specifications of the \texttt{NNFC} and \texttt{CNFC} functions in CPS are similar to the specification of the \texttt{Impl\_Free} function referred above (Appendix \ref{appendix:provacps}). However for the \texttt{CNFC} function it is necessary to prove its pre-conditions. In particular, it is necessary to prove that the input formula is in NNF.

A proof obligation is generated regarding the validity of the pre-condition whenever a recursive
call is made within a continuation. In order to prove such a proof obligation, we need to
specify the nature of the continuation arguments. Thus, we encapsulate the
\texttt{wf\_negations\_of\_literals} predicate into a new type (an invariant type):

\begin{why3}
type nnfc_type = {
    nnfc_formula : formula_wi
} invariant { wf_negations_of_literals nnfc_formula }
  by{ nnfc_formula = FConst_wi true }
\end{why3}
Since the return type of the function is changed, the proof of the post-conditions now involves the comparison of two invariant types, which raises some interesting %difficulties
challenges.
%no entanto como agora o tipo de saída das funções é alterado, a prova das pós-condições implica a comparação de dois tipos invariante, o que levou a algumas dificuldades. %reescrever com frases mais pequenas e introduzir o tipo invariante.

%%%%%%%%%%%%%%%%%%%%%%%%%%%%%%%%%%%%%%%%%%%%%%%%%%%%%%%%%%%%%%%%%%%%%%%%%%%%%%%%%
%%%%%%%%%%%%%%%%%%%%%%%%%%%%%%%%%%%%%%%%%%%%%%%%%%%%%%%%%%%%%%%%%%%%%%%%%%%%%%%%%
%%%%%%%%%%%%%%%%%%%%%%%%%%%%%%%%%%%%%%%%%%%%%%%%%%%%%%%%%%%%%%%%%%%%%%%%%%%%%%%%%
 
\myparagraphbf{Difficulties preventing a proof}
Comparing two invariant types involves providing them a witness, \emph{i.e.}, values with the concerned type; only then it is possible to prove that two values of the same type respect the invariant. However as the invariant type in Why3 is an opaque type, having only access to its projections, it is not possible to construct an inhabitant of this type in the logic, thus making it impossible to compare them. This lemma %can easily 
translates such a behavior:

%Tentando ajudar o Why3 nesta comparação foi criado um lema de igualdade que diz que se o campo fórmula de dois tipos invariantes forem iguais então os records são iguais:

\begin{why3}
lemma types: forall x y. x.cnfc_formula = y.cnfc_formula -> x = y
\end{why3}
It is not possible to prove this lemma because having only access to record projections can not ensure that, in this case, the field \texttt{cnfc\_formula} is the only field of this \textit{record} type.

Given this limitation of Why3 \cite{gitlab}, which in this case precludes the proof of the post-condition, we have tried to compare the formula of each type with an extensional equality predicate (\texttt{==}) and use this predicate as post-condition instead of polymorphic structural equality (\texttt{=}).

\begin{why3}
predicate (==) (t1 t2: cnfc_type) = t1.cnfc_formula = t2.cnfc_formula
\end{why3}
Even with extensional equality, it was not possible to complete the proof. This is due to the fact that for the base cases, given the application to the continuation, we always come across with comparison of records and in the other cases it is not possible to specify the functions of continuation in the recursive calls. This lack of success led to the search for other approaches that would, eventually, achieve the same advantages as the CPS transformation.

\myparagraph{What is the problem with CPS?} The transformation in CPS always adds a function as an argument, thus passing to a higher order function. Since Why3 is a platform that, for reasons of decidability, operates on a first-order language, the solution is to "go back" to first order. The defunctionalization technique emerged as a possible approach.

%% file: English/desfuncionalizationEN.tex
\section{Defunctionalization}

Defunctionalization is a program transformation technique to convert high-order programs into first-order ones \cite{reynolds1972definitional}.
%preciso frase introdutória

\myparagraphbf{Transformation process}
A defunctionalization consists of a "mechanical" transformation in two steps: \ifshortversion representação de primeira ordem das continuações da função e substituição das continuações por esta nova representação; introdução de uma função \texttt{apply} que substitui as aplicações à continuação no programa original. \else
\begin{enumerate}
    \item Get a first order representation of the function continuations and replace the continuations with this new representation.
    \item Generate a new a function \texttt{apply} which replaces the applications of functions
    in the original program.
\end{enumerate}
\fi

\noindent Applying this process to the \texttt{Impl\_Free} function in CPS lead us to represent the function continuations in first-order:

\begin{why3}
type impl_kont =
  | KImpl_Id
  | KImpl_Neg impl_kont formula
  | KImpl_OrLeft formula impl_kont
  | KImpl_OrRight impl_kont formula_wi
  | KImpl_AndLeft formula impl_kont
  | KImpl_AndRight impl_kont formula_wi
  | KImpl_ImplLeft formula impl_kont
  | KImpl_ImplRight impl_kont formula_wi
\end{why3}
The constructor \texttt{KImpl\_id} represents the identity function, the constructor
\texttt{KImpl\_Neg} represents the continuation of the case of the constructor
\texttt{FNeg\_wi}. As the remaining cases contain two continuation functions, two constructors
are created, one \texttt{left} and one \texttt{right}. We chose to use the \texttt{left} and
\texttt{right} nomenclatures because this represents the natural order of the formula in the
abstract syntax tree.

We then replace the continuations with the new representation of the continuations:

\begin{why3}
let rec impl_free_desf_cps (phi: formula) (k: impl_kont) : formula_wi
= match phi with
  | FNeg phi1 -> impl_free_desf_cps phi1 (KImpl_Neg k phi1)
  | FOr phi1 phi2 -> impl_free_desf_cps phi1 (KImpl_OrLeft phi2 k)
  | FAnd phi1 phi2 -> impl_free_desf_cps phi1 (KImpl_AndLeft phi2 k)
  | FImpl phi1 phi2 -> impl_free_desf_cps phi1 (KImpl_ImplLeft phi2 k)
  ...
end
\end{why3}
The next step is to introduce an \texttt{apply} function, and replace the applications to the continuation:

\begin{why3}
with impl_apply (phi: formula_wi) (k: impl_kont) : formula_wi
= match k with
  | KImpl_Id -> phi
  | KImpl_Neg k phi1 -> impl_apply (FNeg_wi phi) k
  | KImpl_OrLeft phi1 k -> impl_free_desf_cps phi1 (KImpl_OrRight k phi)
  | KImpl_OrRight k phi2 -> impl_apply (FOr_wi phi2 phi) k
  | KImpl_AndLeft p k -> impl_free_desf_cps p (KImpl_AndRight k phi)
  | KImpl_AndRight k phi2 -> impl_apply (FAnd_wi phi2 phi) k
  | KImpl_ImplLeft phi1 k -> impl_free_desf_cps phi1 (KImpl_ImplRight k phi)
  | KImpl_ImplRight k phi2-> impl_apply (FOr_wi (FNeg_wi phi2) phi) k
end

let rec impl_free_desf_cps (phi: formula) (k: impl_kont) : formula_wi
= match phi with
  ...
  | FConst phi -> impl_apply (FConst_wi phi) k
  | FVar phi -> impl_apply (FVar_wi phi) k
end
\end{why3}
The result of the application of the defunctionalization transformation to the remaining functions of the T algorithm in CPS is in Appendix~\ref{appendix:desfun}.

\myparagraphbf{Proof of correctness}
The defunctionalized program specification is the same as the original program. However, given the existence of an additional function generated by the defunctionalization process (the \texttt{apply} function), a specification must be provided. Since the \texttt{apply} function simulates the application of a function to its argument, the only specification we can give it is that its post-condition is the post-condition of the function \texttt{k} \cite{pereira2019desfuncionalizar}.

To be able to use the direct-style functions as a specification, we have created a \texttt{post} predicate that gathers the post-conditions of the direct-style function. As for the \texttt{apply} function, such a predicate performs case analysis on the continuation type and for each constructor, we copy the post-condition present in the corresponding abstraction \cite{pereira2019desfuncionalizar}. For instance, for the \texttt{Impl\_Free} function, we provide the following specification

\begin{why3}
let rec impl_free_desf_cps (phi: formula) (k: impl_kont) : formula_wi
  ensures { impl_post k (impl_free phi) result }
 = ...

with impl_apply (phi: formula_wi) (k: impl_kont) : formula_wi
  ensures { impl_post k phi result }
 = ...
\end{why3}
where the \text{impl\_post} predicate is:

\begin{why3}
predicate impl_post (k: impl_kont) (phi result: formula_wi)
= match k with
  | KImpl_Id -> let x = phi in x = result
  | KImpl_Neg k phi1 -> let neg = phi in impl_post k (FNeg_wi phi) result
  | KImpl_OrLeft phi1 k -> let hl = phi in impl_post k (FOr_wi phi
    (impl_free phi1)) result
  | KImpl_OrRight k phi2 -> let hr = phi in impl_post k (FOr_wi phi2 hr) result
  | KImpl_AndLeft phi1 k -> let hl = phi in impl_post k (FAnd_wi phi
    (impl_free phi1)) result
  | KImpl_AndRight k phi2 -> let hr = phi in impl_post k 
    (FAnd_wi phi2 hr) result
  | KImpl_ImplLeft phi1 k -> let hl = phi in impl_post k (FOr_wi 
    (FNeg_wi phi) (impl_free phi1)) result
  | KImpl_ImplRight k phi2-> let hr = phi in impl_post k (FOr_wi 
    (FNeg_wi phi2) hr) result
end
\end{why3}
The proof of the post-conditions of the \texttt{NNFC} and \texttt{CNFC} defunctionalized functions is similar to the proof of the \texttt{Impl\_Free} function (Appendix \ref{appendix:provadesfun}). However, similar to the CPS proof, for the \texttt{CNFC} function, we have to prove its pre-conditions. For this we have created the invariant type \texttt{wf\_cnfc\_kont} with the well-formulated predicate \texttt{wf\_cnfc\_kont} as invariant:

\begin{why3}
type wf_cnfc_kont = {
    cnfc_k: cnfc_kont;
} invariant { wf_cnfc_kont cnfc_k }
  by { cnfc_k = KCnfc_Id }
\end{why3}
Note that in this well-formulated predicate we just want to ensure the CNF for the formulae that
are already converted. Given that the formulae are only converted in the right continuation,
these and only these feature the \texttt{wf\_conjunctions\_of\_disjunctions} predicate:

%A prova da função \texttt{CNFC} desfuncionalizada é semelhante à prova da função \texttt{Distr} desfuncionalizad; as pós-condições são provadas também com um predicado \texttt{post} como especificação e as pré-condições provadas com um tipo invariante e um predicado de boa formação.

\begin{why3}
predicate wf_cnfc_kont (phi: cnfc_kont) 
= match phi with
  | KCnfc_Id -> true
  | KCnfc_OrLeft phi k -> wf_negations_of_literals phi /\ wf_cnfc_kont k
  | KCnfc_OrRight k phi -> wf_negations_of_literals phi /\ wf_conjunctions_of_disjunctions phi /\ wf_cnfc_kont k
  | KCnfc_AndLeft phi k -> wf_negations_of_literals phi /\ wf_cnfc_kont k
  | KCnfc_AndRight k phi -> wf_negations_of_literals phi /\ wf_conjunctions_of_disjunctions phi /\ wf_cnfc_kont k
end
\end{why3}
Lastly, the proof of the T function turns out to be similar to the direct-style proof referenced in Page \pageref{subsec:provat}:

\begin{why3}
let t (phi: formula) : formula_wi
  ensures { forall v. eval v phi = eval_wi v result }
  ensures { wf_negations_of_literals result }
  ensures { wf_conjunctions_of_disjunctions result }
= cnfc_desf_main(nnfc_desf_main(impl_desf_main phi))
\end{why3}

\begin{figure}
    \centering
    \includegraphics[width=\textwidth]{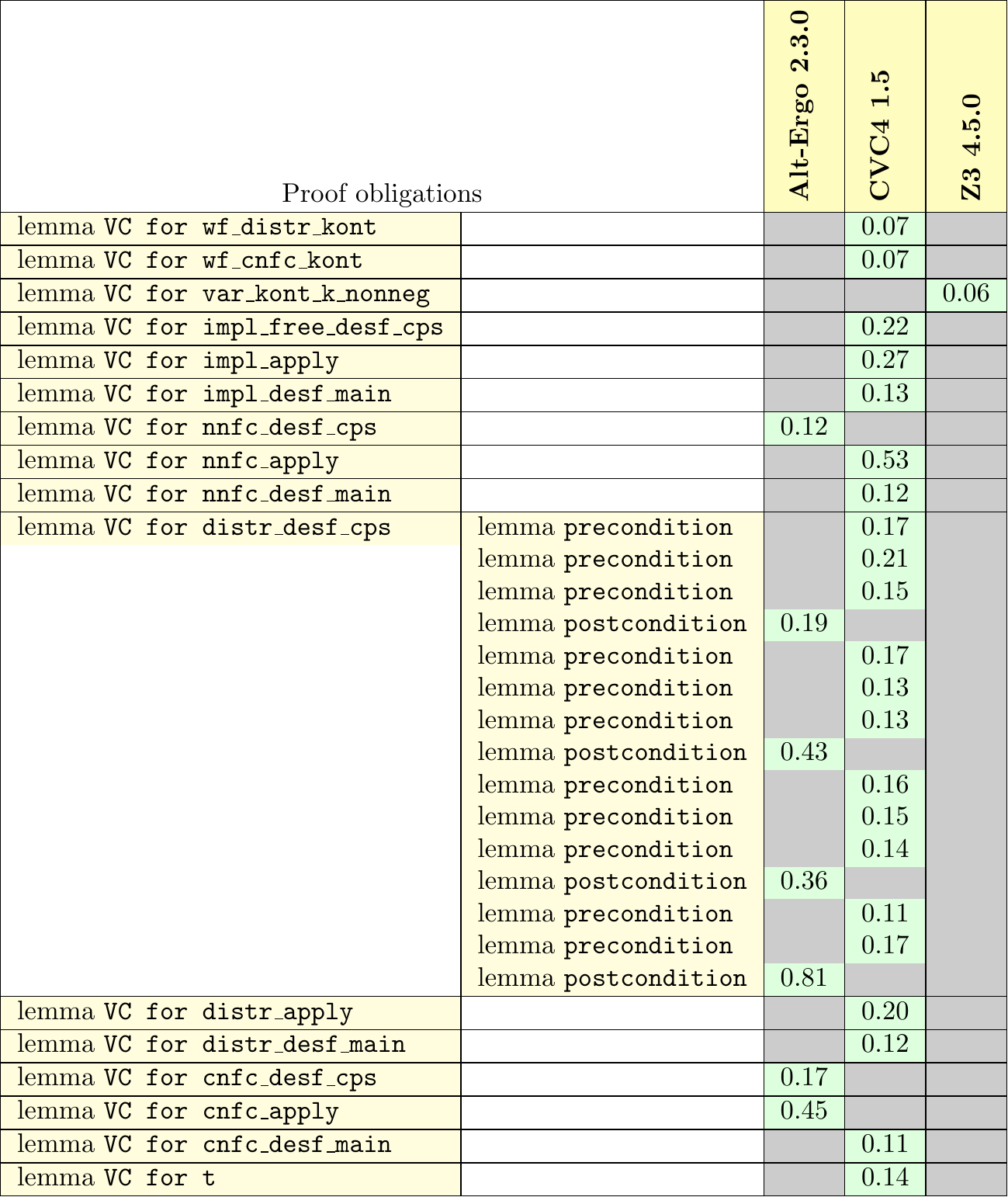} 
    \caption{Statistical result of defunctionalization proof obligations}
    \label{fig:my_label}
\end{figure}

\myparagraphbf{Results}
The proof of correctness of the defunctionalized version of the T algorithm is naturally processed by Why3, with each proof objective being proved in less than one second as %we can see
shown in Figure \ref{fig:my_label}.
%\newpage

%Passar para secção (Conclusao e trabalho futuro) (apendice as duas tabelas)

%% file: English/conclusionEN.tex
\section{Conclusion}
Functional languages such as OCaml allow close implementations of mathematical definitions without
sacrificing clarity and rigor. These make them adequate to be pedagogically used as an aid to the
study and understanding of algorithms.

In this article, we present a proof of concept: the implementation and correctness proof %of
of the algorithm to convert propositional formulae to the Conjunctive Normal Form. Proof of the two strands of the algorithm - direct style and defunctionalized - were achieved naturally by Why3, making successful the proof of concept of formally verified implementations of Computational Logic algorithms.

%In the future, w
We intend to implement the step-by-step execution, through an explicit stack structure, since each function call returns a function (continuation) that can be used as a block, thus allowing to stop and return the execution. We also intend to apply this approach to other Computational Logic algorithms, such as the Horn \cite{hornalgorithm} algorithm.
\newpage
%{\centering \includegraphics[width=\textwidth]{Desfunctionalization.pdf} \par}

%% file: abstract.tex
\begin{abstract}

Neste artigo apresenta-se uma abordagem à implementação formalmente verificada de algoritmos clássicos de Lógica Computacional. Escolhe-se como ferramenta de prova a plataforma Why3 que permite implementações próximas das definições matemáticas, assim como um elevado grau de automação no processo de verificação. Como prova de conceito, utiliza-se o algoritmo de conversão de fórmulas proposicionais para forma normal conjuntiva. Aplica-se a proposta sobre duas variantes deste algoritmo: uma em estilo direto e outra com uma estrutura de pilha explícita no código. Sendo ambas as versões de primeira ordem, o Why3 processa as provas naturalmente.

%Neste artigo implementamos em WhyML o algoritmo que converte fórmulas proposicionais para a forma normal conjuntiva e mostramos a prova de correção do algoritmo; para suportar no futuro a execução passo-a-passo fazemos também uma implementação CPS, as dificuldades que nos deparamos na prova obrigou-nos a passar para a desfuncionalização.

%\keywords{FNC  \and WhyML \and CPS \and Desfuncionalização \and Prova de correção}
\end{abstract}

%% file: introducao.tex
\section{Introdução}
\label{sec:motivation}
\vspace{-0.5em}
\textit{\textbf{Motivação.}} Unidades curriculares centrais em Engenharia Informática, como por exemplo Lógica Computacional, têm como objetivo apresentar conteúdo fundamental para a formação dos estudantes. Para fortalecer a ligação do conteúdo abordado nestas unidades curriculares com a prática de desenvolvimento de código correto, é relevante relacionar o conteúdo matemático a implantações corretas, claras e executáveis.

Este artigo insere-se no projeto \texttt{FACTOR}~\cite{factor}, que visa promover o uso do OCaml~\cite{leroy2014ocaml} e de práticas de desenvolvimento de código correto na comunidade académica de expressão portuguesa. Concretamente, o projeto tem como objetivos a implementação funcional de algoritmos clássicos de Lógica Computacional e Linguagens Formais, a realização de provas de correção das mesmas e a execução passo-a-passo para os ajudar a compreender a partir de exemplos.

O algoritmo de conversão de fórmulas proposicionais para Forma Normal Conjuntiva (FNC)\footnote{Uma fórmula está em FNC se é uma conjunção de cláusulas, onde uma cláusula é uma disjunção de literais, sendo um literal um símbolo proposicional ou a sua negação.} é frequentemente apresentado formalmente, com definições matemáticas rigorosas que, por vezes, são difíceis de ler \cite{inttologicenderton,logichamilton,mathlogicmendelson}, ou informalmente, destinados à Ciência da Computação, mas com definições textuais em pseudo-código não executável \cite{mathlogicbenari,huth2004logic}. A implementação de algoritmos desta natureza é uma peça fundamental para a aprendizagem e compreensão dos mesmos. Linguagens como o OCaml permitem implementações muito próximas das definições matemáticas, ajudando o estudo por serem executáveis. Além disso as prova de correção são mais simples do que as das implementações imperativas.

\myparagraphbf{Contribuições} Como prova de conceito, faz-se a implementação e prova de correção do algoritmo referido em Why3 \cite{filliatre2013why3,why3}, uma plataforma para verificação dedutiva de programas. Why3 fornece uma linguagem de primeira ordem com tipos polimórficos, \textit{pattern matching} e predicados indutivos, chamada WhyML, oferecendo ainda um mecanismo de extração de código OCaml certificado e suporte para provadores de teoremas de terceiros.

Para no futuro suportar a execução passo-a-passo do algoritmo, uma importante funcionalidade para ajudar os estudantes a perceber as definições, implementa-se também uma versão em \textit{Continuation-Passing Style} (CPS) \cite{sabry1993reasoning} e via desfuncionalização obtém-se um avaliador, uma versão próxima a uma máquina abstrata de primeira ordem \cite{danvy2003functionalAM}. Devido ao limitado suporte do Why3 à ordem superior não foi possível fechar a prova de correção. Esta limitação levou ao desenvolvimento de uma implementação com estrutura de pilha explicita no código, mas desta vez em primeira ordem, implementação que resultou de uma transformação mecânica a partir da versão CPS. Esta versão foi naturalmente provada correta pelo Why3.

Em suma, este artigo apresenta material pedagógico de apoio ao ensino de algoritmos clássicos de Lógica Computacional, nomeadamente, duas implementações, formalmente verificadas em Why3, a partir de uma apresentação como função recursiva do \alglong: a primeira em estilo direto e a segunda com estrutura de pilha explicita no código.

O código Why3 apresentado neste artigo pode ser encontrado no repositório público \url{https://bitbucket.org/laforetbarroso/cnfwhy3}.

%% file: descricao.tex
\newcommand\myeq{\mathrel{\stackrel{\makebox[0pt]{\mbox{\normalfont\tiny abv}}}{=}}}
\vspace{-0.5em}
\section{Apresentação funcional do algoritmo}
\label{sec:algdesc}
\vspace{-0.5em}

\mypg{Descrição} Designa-se T ao algoritmo que converte qualquer fórmula de Lógica Proposicional para FNC. Uma fórmula proposicional~$\phi$ é um elemento do conjunto $G_p$, definido como: 
$
G_p \triangleq \ \phi ::= T \ | \ p \ | \ \neg \phi \ | \ \phi \wedge \phi \ | \ \phi \vee \phi \ | \ \phi \rightarrow \phi
$. 

A~função~T produz uma formula sem o conectivo de implicação, portanto define-se um conjunto $H_p$ como um sub-conjunto de $G_p$ sem implicações. Tem-se então que T: $G_p \rightarrow H_p$, onde:
\begin{center}
T($\phi$) = CNFC(NNFC(Impl\_Free($\phi$))) 
\end{center}

\noindent O algoritmo compõe três funções: {\implfree} responsável por eliminar as implicações; {\nnfc} responsável pela conversão para Forma Normal de Negação (FNN)\footnote{Uma fórmula está na FNN, se só as suas sub-fórmulas que são literais estão negadas.}; {\cnfc} responsável pela conversão de FNN para FNC.

\myparagraphbf{Implementação dos conjuntos}
\label{subsec:implalg}
Para representar o conjunto $G_p$ define-se o tipo \texttt{formula} que declara variáveis (\texttt{FVar}), constantes (\texttt{FConst}), conjunções (\texttt{FAnd}), disjunções (\texttt{FOr}), implicações (\texttt{FImpl}) e negações (\texttt{FNeg}):

\begin{why3}
type formula =
  | FVar ident
  | FConst bool
  | FAnd formula formula
  | FOr formula formula
  | FImpl formula formula
  | FNeg formula
\end{why3}
%Estrategicamente escolhemos internalizar a ausência de implicações no tipo \texttt{formula\_wi}.
Para representar o conjunto $H_p$ define-se o tipo \texttt{formula\_wi}, semelhante ao anterior mas sem o construtor de implicação. 

%Apresentamos de seguida a implementação das três funções auxiliares:

\myparagraphbf{Implementação das funções} A função \texttt{Impl\_Free} elimina todas as implicações. É definida recursivamente nos casos do tipo \texttt{formula} e homomórfica, excepto no caso da implicação, onde utiliza a lei de Lógica Proposicional 
$\mathtt{A} \rightarrow \mathtt{B} \equiv \neg \mathtt{A} \vee \mathtt{B}$.
A~implementação da função converte os construtores do tipo \texttt{formula} 
para os do tipo \texttt{formula\_wi} e efetua chamadas recursivas sobre os argumentos:
\vspace{-0.3em}
\begin{why3}
let rec impl_free (phi: formula) : formula_wi
= match phi with
  | FNeg phi1 -> FNeg_wi (impl_free phi1)
  | FOr phi1 phi2 -> FOr_wi (impl_free phi1) (impl_free phi2)
  | FAnd phi1 phi2 -> FAnd_wi (impl_free phi1) (impl_free phi2)
  | FImpl phi1 phi2 -> FOr_wi (FNeg_wi (impl_free phi1)) (impl_free phi2)
  | FConst phi -> FConst_wi phi
  | FVar phi -> FVar_wi phi
end
\end{why3}
\vspace{-0.3em}
A função {\nnfc} converte a fórmula para a FNN. É definida recursivamente em combinações de construtores: as duplas negações são eliminadas aplicando a lei de Lógica Proposicional
\texttt{$\neg \neg$A $\equiv$ A} e, usando as leis de De Morgan, as negações de conjunções passam a disjunções de negações e as negações de disjunções passam a conjunções de negações. O código da função é o seguinte:
\vspace{-0.3em}
\begin{why3}
let rec nnfc (phi: formula_wi) : formula_wi
= match phi with
  | FNeg_wi (FNeg_wi phi1) -> nnfc phi1
  | FNeg_wi (FAnd_wi phi1 phi2) -> FOr_wi (nnfc (FNeg_wi phi1)) 
    (nnfc (FNeg_wi phi2))
  | FNeg_wi (FOr_wi phi1 phi2) -> FAnd_wi (nnfc (FNeg_wi phi1)) 
    (nnfc (FNeg_wi phi2))
  | FOr_wi phi1 phi2 -> FOr_wi (nnfc phi1) (nnfc phi2)
  | FAnd_wi phi1 phi2 -> FAnd_wi (nnfc phi1) (nnfc phi2)
  | phi -> phi
end
\end{why3}
\vspace{-0.3em}
A função {\cnfc} converte uma fórmula em FNN para FNC, sendo homomórfica excepto no caso da disjunção, onde efetua a distribuição da disjunção pela conjunção chamando a função auxiliar \texttt{distr}.
\vspace{-0.3em}
\begin{why3}
let rec cnfc (phi: formula_wi) : formula_wi
= match phi with
  | FOr_wi phi1 phi2 -> distr (cnfc phi1) (cnfc phi2)
  | FAnd_wi phi1 phi2 -> FAnd_wi (cnfc phi1) (cnfc phi2)
  | phi -> phi
end
\end{why3}
\vspace{-0.3em}
A função \texttt{distr} por sua vez tira partido da lei de Lógica Proposicional:
$$\mathtt{A} \vee (\mathtt{B} \wedge \mathtt{C}) \equiv (\mathtt{A} \vee \mathtt{B}) \wedge (\mathtt{A} \vee \mathtt{C})$$

\noindent O código da função \texttt{distr} é o seguinte:
\vspace{-0.3em}
\begin{why3} 
let rec distr (phi1 phi2: formula_wi) : formula_wi
= match phi1, phi2 with
  | FAnd_wi phi11 phi12, phi2 -> FAnd_wi (distr phi11 phi2) 
    (distr phi12 phi2)
  | phi1, FAnd_wi phi21 phi22 -> FAnd_wi (distr phi1 phi21) 
    (distr phi1 phi22)
  | phi1,phi2 -> FOr_wi phi1 phi2
end
\end{why3}
\vspace{-0.3em}
Finalmente, o código da função T que compõe todas estas funções é o seguinte:
\vspace{-0.3em}
\begin{why3}
let t (phi: formula) : formula_wi 
= cnfc(nnfc(impl_free phi))
\end{why3}
\vspace{-0.3em}
A partir desta implementação é possível a extração de código OCaml.
Em ambas as implementações ressalta a semelhança com as definições matemáticas, demonstrando assim que o OCaml é uma linguagem adequada para a apresentação destes algoritmos, providenciando definições executáveis sem sacrifício de rigor ou clareza.

%% file: criterios.tex
\vspace{-0.5em}
\section{Como obter a correção}%Correção da Implementação}
\label{sec:corrcrite}
\vspace{-0.5em}
O algoritmo T, como apresentado anteriormente, é uma composição de três funções, sendo a correção do algoritmo resultado dos critérios de correção de cada uma dessas três funções. 

\myparagraphbf{Critérios}
\label{sec:corrcrite} %reformular e introduzir pós e pré-condições, perguntar.
Para todas as funções o critério de correção básico é que o seu resultado deve ser uma fórmula equivalente à formula argumento. Além disso requer-se que: \ifshortversion
 o resultado da função \texttt{Impl\_Free} não deve conter conetivos de implicação; a fórmula de entrada e resultado da função \texttt{NNFC} não devem conter conetivos de implicação e o resultado tem que estar em FNN; a fórmula de entrada e resultado da função \texttt{CNFC} não devem conter conetivos de implicação e tem que estar em FNN e o resultado tem que estar em FNC.

\else
\paragraph{Impl\_Free}

\begin{itemize}
	\item O resultado não deve conter conetivos de implicação.
\end{itemize}

\paragraph{NNFC}

\begin{itemize}
	\item A fórmula de entrada e resultado não deve conter conetivos de implicação.
	\item O resultado tem que estar na forma normal de negação.
\end{itemize}

\paragraph{CNFC}

\begin{itemize}
	\item A fórmula de entrada e resultado não deve conter conetivos de implicação.
	\item A fórmula de entrada e resultado tem que estar na forma normal de negação.
	\item O resultado tem que estar na forma normal conjuntiva.
\end{itemize}
\fi
De notar que os critérios de correção de uma função são propagados para as funções seguintes, garantindo que uma função não viola as pós-condições já assegurados pelas funções previamente executadas.

\myparagraphbf{Semântica das fórmulas}
Como o critério básico de correção é a equivalência de fórmulas, é preciso uma função para as avaliar (ou seja, de valoração):
\vspace{-0.3em}
\begin{why3}
type valuation = ident -> bool

function eval (v: valuation) (f: formula) : bool
= match f with
  | FVar x      -> v x
  | FConst b    -> b
  | FAnd f1 f2  -> eval v f1 /\ eval v f2
  | FOr f1 f2   -> eval v f1 || eval v f2
  | FImpl f1 f2 -> eval v f1 -> eval v f2
  | FNeg f      -> not (eval v f)
end
\end{why3}
Esta função recebe um argumento de tipo \texttt{valuation} que atribui um valor do tipo \texttt{bool}\footnote{bool é o tipo booleano do WhyML} a cada variável da fórmula, recebe a \texttt{fórmula} a ser valorada e retorna um valor do tipo \texttt{bool}. Para os construtores base \texttt{FVar} e \texttt{FConst}, apenas é retornado o valor booleano da variável e o valor da constante, respetivamente. Para os restantes casos construtores são valoradas recursivamente as fórmulas associadas e o resultado traduzido para a operação booleana correspondente do WhyML. A função de valoração para o tipo de fórmulas \texttt{formula\_wi} é semelhante.

%% file: estilodireto.tex
\vspace{-0.5em}
\section{Prova de correção}
\vspace{-0.5em}
A prova da correção da implementação consiste em mostrar que cada função respeita os critérios de correção definidos na secção anterior.

\myparagraphbf{Palavras-chave} Em WhyML as palavras-chaves \texttt{ensures} correspondem à indicação de pós-condições, consequentemente, à prova de correção parcial. A terminação e prova de correção total são asseguradas com a palavra-chave \texttt{variant}.

\myparagraphbf{Correção da função Impl\_Free} A ausência de conetivos de implicação é assegurado pelo tipo de retorno da função (\texttt{formula\_wi}); a equivalência das fórmulas é assegurada usando as funções de valoração de fórmulas e usa-se a fórmula de entrada como medida para garantir a terminação.
\vspace{-0.3em}
\begin{why3}
let rec function impl_free (phi: formula) : formula_wi
  ensures { forall v. eval v phi = eval_wi v result }
  variant { phi }
= ...
\end{why3}
\vspace{-0.3em}
\myparagraphbf{Correção da função NNFC} %talvez mudar FNN para forma normal de negação
A ausência de conetivos de implicação nas fórmulas de entrada e saída é assegurada pelo tipo \texttt{formula\_wi}. Para provar que o resultado está na FNN, submete-se à prova o predicado de boa formação  \texttt{wf\_negations\_of\_literals}. Este estabelece que as sub-fórmulas do construtor \texttt{FNeg\_wi} não podem conter construtores \texttt{FOr\_wi}, \texttt{FAnd\_wi} ou \texttt{FNeg\_wi}:
\vspace{-0.3em}
\begin{why3}
predicate wf_negations_of_literals (f: formula_wi)
= match f with
  | FNeg_wi f -> (forall f1 f2. f <> FOr_wi f1 f2 /\ f <> FAnd_wi f1 f2 /\ f <> FNeg_wi f1) /\ wf_negations_of_literals f
  | FOr_wi f1 f2 | FAnd_wi f1 f2 -> wf_negations_of_literals f1 /\ wf_negations_of_literals f2
  | FVar_wi _ -> true
  | FConst_wi _ -> true
end
\end{why3}
\vspace{-0.3em}
Nesta prova não é possível usar a própria fórmula como medida de terminação visto que no caso da distribuição da negação pela conjunção ou disjunção são adicionados construtores à cabeça, impossibilitando o critério indutivo estrutural. Criou-se então, uma função que conta o número de construtores de uma fórmula. No entanto, para ser possível usá-la como medida de terminação é preciso assegurar, com um lema, que o número de construtores nunca é negativo.

Com o predicado e medida de terminação definidos é possível fechar a prova de correção da função \texttt{NNFC}, sendo o código submetido à prova o seguinte:
\begin{why3}
let rec nnfc (phi: formula_wi) : formula_wi
  ensures { forall v. eval_wi v phi = eval_wi v result }
  ensures { wf_negations_of_literals result }
  variant { size phi }
= ...
\end{why3}
\vspace{-0.3em}
\myparagraphbf{Prova da função CNFC}
O critério básico de equivalência é mais uma vez assegurado usando a função de valoração de fórmulas. Para assegurar que uma determinada fórmula está na FNC introduzem-se os predicados de boa formação \texttt{wf\_conjunctions\_of\_disjunctions} e \texttt{wf\_disjunctions}. Estes garantem que após uma disjunção não há nenhuma conjunção:
\vspace{-0.3em}
\begin{why3}
predicate wf_conjunctions_of_disjunctions (f: formula_wi)
= match f with
  | FAnd_wi f1 f2 -> wf_conjunctions_of_disjunctions f1 /\ wf_conjunctions_of_disjunctions f2
  | FOr_wi f1 f2 -> wf_disjunctions f1 /\ wf_disjunctions f2
  | FConst_wi _ -> true
  | FVar_wi _ -> true
  | FNeg_wi f1 -> wf_conjunctions_of_disjunctions f1
end

predicate wf_disjunctions (f: formula_wi)
= match f with
  | FAnd_wi _ _ -> false
  | FOr_wi f1 f2 -> wf_disjunctions f1 /\ wf_disjunctions f2
  | FConst_wi _ -> true
  | FVar_wi _ -> true
  | FNeg_wi f1 -> wf_disjunctions f1
end
\end{why3}
\vspace{-0.3em}
Finalmente, adicionam-se os predicados \texttt{wf\_conjunctions\_of\_disjunctions} e \texttt{wf\_negations\_of\_literals} às pós-condições para assegurar que o resultado está na FNN e FNC, respetivamente; para assegurar que a fórmula de entrada está na FNN, adiciona-se também o predicado \texttt{wf\_negations\_of\_literals} às pré-condições. O código submetido à prova de correção é o seguinte:
\vspace{-0.3em}
\begin{why3}
let rec cnfc (phi: formula_wi)
  requires{ wf_negations_of_literals phi }
  ensures{ forall v. eval_wi v phi = eval_wi v result }
  ensures{ wf_negations_of_literals result }
  ensures{ wf_conjunctions_of_disjunctions result } 
  variant { phi }
= ...
\end{why3}
\vspace{-0.3em}
Sendo \texttt{distr} uma função auxiliar da função \texttt{CNFC}, é preciso também provar a correção da mesma. Nesta função é necessário garantir os mesmos critérios da função \texttt{CNFC}, mas por se tratar da distribuição das disjunções pelas conjunções, é preciso adicionalmente assegurar que as fórmulas de entrada estão na FNC, o que se obtém adicionando os predicados \texttt{wf\_conjunctions\_of\_disjunctions} e \texttt{wf\_negations\_of\_literals} às pré-condições:
\vspace{-0.3em}
\begin{why3}
let rec distr (phi1 phi2: formula_wi)
  requires{ wf_negations_of_literals phi1 }
  requires{ wf_negations_of_literals phi2 }
  requires{ wf_conjunctions_of_disjunctions phi1 } 
  requires{ wf_conjunctions_of_disjunctions phi2 }
  ensures { forall v. eval_wi v (FOr_wi phi1 phi2) = eval_wi v result }
  ensures { wf_negations_of_literals result }
  ensures { wf_conjunctions_of_disjunctions result }
  variant { size phi1 + size phi2 }
= ...
\end{why3}
\vspace{-0.3em}
No entanto, não se consegue provar que uma disjunção de duas fórmulas na FNC é efetivamente uma fórmula na FNC\ifshortversion 
\else
. A seguinte imagem representa o caso; a verde está a informação que o Why3 tem e a amarelo simbolizado a prova do predicado para o caso em especifico:

{\centering \includegraphics[width=1\textwidth]{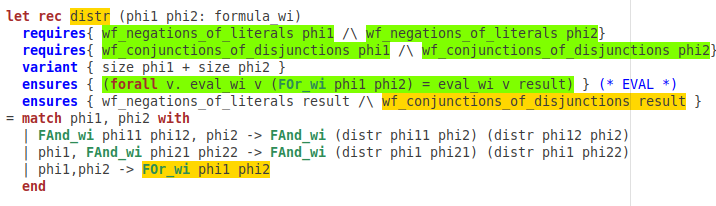} \par}
\fi
, isto porque é necessário assegurar que numa disjunção de duas fórmulas na FNC, as fórmulas não contêm o construtor \texttt{FAnd\_wi}. Para tal, reforça-se a prova com um lema auxiliar:
\vspace{-0.3em}
\begin{why3}
lemma aux: forall x. wf_conjunctions_of_disjunctions x /\
wf_negations_of_literals x /\ not (exists f1 f2. x = FAnd_wi f1 f2) ->
wf_disjunctions x
\end{why3}
\vspace{-0.3em}
\myparagraphbf{Prova da função T}
\label{subsec:provat}
Com as provas de correção de cada uma das três funções efetuadas, pode-se agora obter a prova de correção da função T. Esta garante todos os critérios assegurados pelas três funções:
\vspace{-0.3em}
%comentário sobre cada ensures
\begin{why3}
let t (phi: formula) : formula_wi
  ensures { wf_negations_of_literals result }
  ensures { wf_conjunctions_of_disjunctions result }
  ensures { forall v. eval v phi = eval_wi v result  }
= cnfc (nnfc (impl_free phi))
\end{why3}
\vspace{-0.3em}
A prova em estilo direto da implementação e especificação -- próxima das definições matemáticas clássicas -- é imediata em Why3, tornando este exercício numa bem sucedida prova de conceito.

%% file: cps.tex
\vspace{-0.5em}
\section{Continuation-Passing Style}
\vspace{-0.5em}
\textit{Continuation-Passing Style} (CPS) é um estilo de programação onde o controlo é passado explicitamente na forma de continuação, evitando assim o \textit{overflow} da pilha se o compilador subjacente optimizar as chamadas recursivas terminais. Com uma estrutura de pilha explicita no código é possível, no futuro, introduzir um mecanismo que permita a execução passo-a-passo das funções.

\myparagraphbf{Processo de transformação para CPS}

A transformação é efetuada de forma mecânica e segue os seguintes passos: dada uma função \texttt{t' $\rightarrow$ t}, adiciona-se um argumento que representará a continuação (uma função do tipo \texttt{t $\rightarrow$ 'a}) e é alterado o tipo de retorno da função para \texttt{'a}; para os casos bases em vez de se retornar os valores desejados, aplicam-se estes valores à função de continuação; para os restantes casos, começa-se por efetuar uma chamada recursiva à função, sendo as continuações criadas com o resto da computação. Por fim, é criada uma função \texttt{main} que chama a função CPS com a função identidade como~continuação.

Aplicando este processo à função \texttt{Impl\_Free}: \ifshortversion adiciona-se um argumento à função do tipo \texttt{formula\_wi $\rightarrow$ 'a} e altera-se o tipo de retorno para \texttt{'a}: \vspace{-0.3em}\begin{why3}
let rec impl_free_cps (phi: formula) (k: formula_wi -> 'a ) : 'a
\end{why3}\vspace{-0.3em}
Os casos bases são então aplicados à função de continuação: \vspace{-0.3em}\begin{why3}
| FConst phi -> k (FConst_wi phi)
| FVar phi -> k (FVar_wi phi)
\end{why3}
\vspace{-0.3em}
Para os restantes casos começa-se com uma chamada recursiva e define-se as continuações: \vspace{-0.3em}\begin{why3}
| FNeg phi1 -> impl_free_cps phi1 (fun con -> k (FNeg_wi con))
| FImpl phi1 phi2 -> impl_free_cps phi1 (fun con -> impl_free_cps phi2 
    (fun con1 -> k (FOr_wi (FNeg_wi con) con1)))
...\end{why3}
\vspace{-0.3em}
Por fim, cria-se uma função \texttt{main} que chama a função CPS com a função identidade como continuação:
\vspace{-0.3em}\begin{why3}
let impl_free_main (phi: formula) : formula_wi
= impl_free_cps phi (fun x -> x)\end{why3}\vspace{-0.3em}
\else
\begin{enumerate}
    \item Adicionar um novo argumento à função do tipo \texttt{formula\_wi $\rightarrow$ 'a} e o tipo de saída é alterado para 'a. \begin{why3}
let rec impl_free_cps (phi: formula) (k: formula_wi -> 'a ) : 'a
\end{why3}
    \item Os casos bases são aplicados à função de continuação. \begin{why3}
    | FConst phi -> k (FConst_wi phi)
    | FVar phi -> k (FVar_wi phi)
\end{why3}
    \item Para os restantes casos começamos com uma chamada recursiva e definimos as continuações. \begin{why3}| FNeg phi1 -> impl_free_cps phi1 (fun con -> k (FNeg_wi con))
    | FOr phi1 phi2 -> impl_free_cps phi1 (fun con -> impl_free_cps phi2 (fun con1 -> k (FOr_wi con con1)))
    | FAnd phi1 phi2 -> impl_free_cps phi1 (fun con -> impl_free_cps phi2 (fun con1 -> k (FAnd_wi con con1)))
    | FImpl phi1 phi2 -> impl_free_cps phi1 (fun con -> impl_free_cps phi2 (fun con1 -> k (FOr_wi (FNeg_wi con) con1)))\end{why3}
    \item Criação de uma função \texttt{main} responsável por chamar a função em CPS com a função identidade como continuação.
    \begin{why3}
let impl_free_main (phi: formula) : formula_wi
  = impl_free_cps phi (fun x -> x)\end{why3}
\end{enumerate}
\fi
A transformação em CPS das restantes funções é obtida de forma semelhante.

\myparagraphbf{Especificação dos critérios de correção}
Um aspecto interessante na prova de correção das funções em CPS é o uso da correspondente função em estilo direto, visto estas serem puras e totais, como própria especificação, ou seja, assegura-se que o resultado é igual ao resultado das funções em estilo direto aplicado à continuação.

Para a função {\implfree} em CPS basta então assegurar que o resultado é igual ao resultado da função {\implfree} em estilo direto aplicado à continuação:

\begin{why3}
let rec impl_free_cps (phi: formula) (k: formula_wi -> 'a ) : 'a
  variant { phi }
  ensures { result = k(impl_free phi) }
= ...
\end{why3}
\vspace{-0.3em}
A especificação da função em estilo direto é depois aplicada à função \texttt{main}:
\vspace{-0.3em}\begin{why3}
let impl_free_main (phi: formula) : formula_wi
  ensures { forall v. eval v phi = eval_wi v result }
= ...
\end{why3}\vspace{-0.3em}
As especificações das funções \texttt{NNFC} e \texttt{CNFC} em CPS são semelhantes à especificação da função \texttt{Impl\_Free}; no entanto, é necessário provar as pré-condições da função \texttt{CNFC}, ou seja, provar que a fórmula de entrada está na FNN. 

Sempre que é efectuada uma chamada recursiva no interior de uma continuação, uma obrigação de prova é gerada respeitante à validade da pré-condição desta chamada. Para provar tal obrigação de prova, é necessário especificar a natureza dos argumentos das continuações. Assim, encapsula-se o predicado \texttt{wf\_negations\_of\_literals} dentro de um novo tipo (tipo invariante):
\vspace{-0.3em}
\begin{why3}
type nnfc_type = {
    nnfc_formula : formula_wi
  } invariant { wf_negations_of_literals nnfc_formula }
    by{ nnfc_formula = FConst_wi true }
\end{why3}
\vspace{-0.3em}
Visto que o tipo de retorno da função é alterado, a prova das pós-condições implica agora a comparação de dois tipos invariante, o que levantou dificuldades.

\myparagraphbf{Dificuldades da prova}
Comparar dois tipos invariante implica dar-lhes uma testemunha, ou seja, valores com o tipo em causa; só assim é possível provar que dois valores do mesmo tipo respeitam o invariante; no entanto como o tipo invariante em Why3 é um tipo opaco, tendo apenas acesso às suas projeções, não é possível a construção de um habitante deste tipo na lógica, impossibilitando assim a sua comparação. Esta dificuldade é facilmente traduzida para um lema:

\begin{why3}
lemma types: forall x y. x.cnfc_formula = y.cnfc_formula -> x = y 
\end{why3}
Não é possível provar este lema porque tendo apenas acesso às projeções do \textit{record} não é possível assegurar que, neste caso, o campo \texttt{cnfc\_formula} é o único campo do \textit{record}. Tendo em conta esta limitação do Why3 \cite{gitlab}, o que neste caso impossibilita a prova da pós-condição, tentou-se apenas comparar a fórmula de cada tipo com um predicado de igualdade extensional (\texttt{==}) e usar este predicado como pós-condição em vez da igualdade estrutural polimórfica (\texttt{=}).

\begin{why3}
predicate (==) (t1 t2: cnfc_type) = t1.cnfc_formula = t2.cnfc_formula
\end{why3}
Mesmo com a igualdade extensional, não foi possível concluir a prova. Isto porque, nos casos bases, devido à aplicação à continuação, acaba-se sempre por deparar com uma comparação de records e nos restantes casos não é possível especificar as funções de continuação nas chamadas recursivas. Esta falta de sucesso levou à procura de outras abordagens que permitissem obter as mesmas vantagens que a transformação CPS.

\myparagraph{Qual o problema com CPS?} A transformação em CPS adiciona sempre uma função como argumento, passando assim a uma função de ordem superior. Sendo o Why3 uma plataforma que por razões de decidibilidade opera sobre uma linguagem de primeira ordem, a solução passa por ``voltar" para a primeira ordem, surgindo então a desfuncionalização como uma possível abordagem.

%% file: desfuncionalizacao.tex
\vspace{-0.5em}
\section{Desfuncionalização}
\vspace{-0.5em}

A desfuncionalização é uma técnica de transformação de programas de ordem superior para programas de primeira ordem \cite{reynolds1972definitional}.

\myparagraphbf{Processo de transformação}
A desfuncionalização consiste numa transforma\-ção ``mecânica'' em dois passos: \ifshortversion representação de primeira ordem das continuações da função e substituição das continuações por esta nova representação; introdução de uma função \texttt{apply} que substitui as aplicações à continuação no programa original. \else
\begin{enumerate}
    \item Representação de primeira ordem das continuações da função e substituição das continuações por esta nova representação.
    \item Introdução de uma função \texttt{apply} que substitui as aplicações à continuação no programa original.
\end{enumerate}
\fi
Aplicando este processo à função \texttt{Impl\_Free} em CPS, representam-se em primeira ordem as continuações da função:
\vspace{-0.3em}
\begin{why3}
type impl_kont =
  | KImpl_Id
  | KImpl_Neg impl_kont formula
  | KImpl_OrLeft formula impl_kont
  | KImpl_OrRight impl_kont formula_wi
  ...
\end{why3}
\vspace{-0.3em}
O construtor \texttt{KImpl\_id} representa a identidade e o construtor \texttt{KImpl\_Neg} representa a continuação do caso do construtor {FNeg\_wi}. Como os restantes construtores contêm duas funções de continuação criam-se dois construtores, um \texttt{left} e um \texttt{right}, representado assim a ordem da formula na árvore sintaxe abstrata.

Depois substituí-se as continuações pela nova representação das continuações, introduz-se uma função \texttt{apply} e substitui-se as aplicações à continuação:
\vspace{-0.3em}
\begin{why3}
let rec impl_free_desf_cps (phi: formula) (k: impl_kont) : formula_wi
= match phi with
  | FNeg phi1 -> impl_free_desf_cps phi1 (KImpl_Neg k phi1)
  | FOr phi1 phi2 -> impl_free_desf_cps phi1 (KImpl_OrLeft phi2 k)
  | FVar phi -> impl_apply (FVar_wi phi) k
  ...
end
  
with impl_apply (phi: formula_wi) (k: impl_kont) : formula_wi
= match k with
  | KImpl_Id -> phi
  | KImpl_Neg k phi1 -> impl_apply (FNeg_wi phi) k
  | KImpl_OrLeft phi1 k -> impl_free_desf_cps phi1 (KImpl_OrRight k phi)
  | KImpl_OrRight k phi2 -> impl_apply (FOr_wi phi2 phi) k
  ...
end
\end{why3}
\vspace{-0.3em}
As transformações das restantes funções são obtidas de forma semelhante.

\myparagraphbf{Prova de Correção}
A especificação do programa desfuncionalizado é a mesma do programa original; no entanto, dada a existência de uma função adicional (a função \texttt{apply} gerada pelo processo de desfuncionalização), é preciso fornecer uma especificação a esta. Sendo a função \texttt{apply} uma simulação da aplicação de uma função ao seu argumento, a única especificação que se pode fornecer é a de que a sua pós-condição é a pós-condição da função \texttt{k} \cite{pereira2019desfuncionalizar}.

Para ser possível o uso das funções em estilo direto como especificação, cria-se um predicado \texttt{post} que reúne as pós-condições da função em estilo direto. Tal como para a função apply, um tal predicado efectua uma filtragem sobre o tipo da continuação e para cada construtor copia-se a pós-condição presente na abstracção correspondente \cite{pereira2019desfuncionalizar}. Por exemplo para a função \texttt{Impl\_Free}:

\begin{why3}
let rec impl_free_desf_cps (phi: formula) (k: impl_kont) : formula_wi
  ensures{impl_post k (impl_free phi) result}
 = ...

with impl_apply (phi: formula_wi) (k: impl_kont) : formula_wi
  ensures{impl_post k phi result}
 = ...
\end{why3}
\vspace{-0.3em}
Sendo o predicado \text{impl\_post} o seguinte:
\vspace{-0.3em}
\begin{why3}
predicate impl_post (k: impl_kont) (phi result: formula_wi)
= match k with
  | KImpl_Id -> let x = phi in x = result
  | KImpl_Neg k phi1 -> let neg = phi in impl_post k (FNeg_wi phi) result
  | KImpl_OrLeft phi1 k -> let hl = phi in impl_post k (FOr_wi phi
    (impl_free phi1)) result
  | KImpl_OrRight k phi2 -> let hr = phi in impl_post k (FOr_wi phi2 hr) result
  ...
end
\end{why3}
\vspace{-0.3em}
A prova das pós-condições das restantes funções desfuncionalizadas é semelhante à da função \texttt{Impl\_Free}. No entanto, à semelhança da prova em CPS, na função \texttt{CNFC} é preciso provar as suas pré-condições. Para tal, cria-se o tipo invariante \texttt{wf\_cnfc\_kont} com o predicado de boa formação \texttt{wf\_cnfc\_kont} como~invariante:

\begin{why3}
type wf_cnfc_kont = {
    cnfc_k: cnfc_kont;
  } invariant { wf_cnfc_kont cnfc_k }
    by { cnfc_k = KCnfc_Id }
\end{why3}
\vspace{-0.3em}
De notar que no predicado de boa formação apenas se quer assegurar a FNC para as fórmulas que já estão convertidas. Tendo em conta que as fórmulas só são convertidas na continuação da direita, apenas estas e só estas contêm o predicado \texttt{wf\_conjunctions\_of\_disjunctions}:

%A prova da função \texttt{CNFC} desfuncionalizada é semelhante à prova da função \texttt{Distr} desfuncionalizad; as pós-condições são provadas também com um predicado \texttt{post} como especificação e as pré-condições provadas com um tipo invariante e um predicado de boa formação.
\vspace{-0.3em}
\begin{why3}
predicate wf_cnfc_kont (phi: cnfc_kont) 
= match phi with
  | KCnfc_Id -> true
  | KCnfc_OrLeft phi k -> wf_negations_of_literals phi /\ wf_cnfc_kont k
  | KCnfc_OrRight k phi -> wf_negations_of_literals phi /\ wf_conjunctions_of_disjunctions phi /\ wf_cnfc_kont k
  ...
end
\end{why3}
\vspace{-0.3em}
Finalmente, a prova da função T acaba por ser semelhante à prova em estilo direto referida na Página \pageref{subsec:provat}.

\myparagraphbf{Resultados}
A prova de correção da versão desfuncionalizada do algoritmo T é processada naturalmente pelo Why3, sendo a prova de cada objetivo de prova realizada em menos de um segundo. O resultado da prova pode ser observado no repositório do projeto.

%Passar para secção (Conclusao e trabalho futuro) (apendice as duas tabelas)

%% file: conclusao.tex
\vspace{-0.5em}
\section{Conclusão}
\vspace{-0.5em}
Linguagens como o OCaml, permitem implementações próximas das definições matemáticas, sem sacrificar clareza e rigor, sendo adequadas ao uso pedagógico de auxílio ao estudo e compreensão de algoritmos.

Neste artigo apresenta-se uma prova de conceito: a implementação e prova de correção da conversão de fórmulas proposicionais para a FNC. A prova das duas vertentes do algoritmo -- estilo direto e desfuncionalizada -- foi conseguida naturalmente pelo Why3, tornando bem sucedida a prova de conceito de implementações formalmente verificadas de algoritmos de Lógica Computacional.

Futuramente, pretende-se efetuar implementações suportando a execução passo-a-passo, através de uma estrutura de pilha explicita no código, visto que cada chamada da função retorna uma função (continuação) que pode ser usada como bloqueio, permitindo assim a paragem e retorno da execução. Pretende-se também aplicar esta abordagem a outros algoritmos de Lógica Computacional, como por exemplo o algoritmo de Horn \cite{hornalgorithm}.

%{\centering \includegraphics[width=\textwidth]{Desfunctionalization.pdf} \par}

%% file: English/appendixEN.tex
\newpage

\section{OCaml code extracted from the WhyML implementation}
\label{appendix:codigoocaml}

\begin{why3}
type ident = string

type formula =
  | FVar of ident
  | FConst of bool
  | FAnd of formula * formula
  | FOr of formula * formula
  | FImpl of formula * formula
  | FNeg of formula

type formula_wi =
  | FVar_wi of ident
  | FConst_wi of bool
  | FAnd_wi of formula_wi * formula_wi
  | FOr_wi of formula_wi * formula_wi
  | FNeg_wi of formula_wi
  
let rec impl_free (phi: formula) : formula_wi =
  begin match phi with
  | FNeg phi1 -> FNeg_wi (impl_free phi1)
  | FOr (phi1, phi2) -> FOr_wi ((impl_free phi1), (impl_free phi2))
  | FAnd (phi1, phi2) -> FAnd_wi ((impl_free phi1), (impl_free phi2))
  | FImpl (phi1, phi2) -> FOr_wi ((FNeg_wi (impl_free phi1)), (impl_free phi2))
  | FConst phi1 -> FConst_wi phi1
  | FVar phi1 -> FVar_wi phi1
  end

let rec nnfc (phi: formula_wi) : formula_wi =
  begin match phi with
  | FNeg_wi (FNeg_wi phi1) -> nnfc phi1
  | FNeg_wi (FAnd_wi (phi1, phi2)) -> FOr_wi ((nnfc (FNeg_wi phi1)), (nnfc (FNeg_wi phi2)))
  | FNeg_wi (FOr_wi (phi1, phi2)) -> FAnd_wi ((nnfc (FNeg_wi phi1)), (nnfc (FNeg_wi phi2)))
  | FOr_wi (phi1, phi2) -> FOr_wi ((nnfc phi1), (nnfc phi2))
  | FAnd_wi (phi1, phi2) -> FAnd_wi ((nnfc phi1), (nnfc phi2))
  | phi1 -> phi1
  end

let rec distr (phi1: formula_wi) (phi2: formula_wi) : formula_wi =
  begin match (phi1, phi2) with
  | (FAnd_wi (phi11, phi12), phi21) ->
    let o = distr phi12 phi21 in let o1 = distr phi11 phi21 in FAnd_wi (o1, o)
  | (phi11, FAnd_wi (phi21, phi22)) ->
    let o = distr phi11 phi22 in let o1 = distr phi11 phi21 in FAnd_wi (o1, o)
  | (phi11, phi21) -> FOr_wi (phi11, phi21)
  end

let rec cnfc (phi: formula_wi) : formula_wi =
  begin match phi with
  | FOr_wi (phi1, phi2) -> let o = cnfc phi2 in let o1 = cnfc phi1 in distr o1 o
  | FAnd_wi (phi1, phi2) ->
    let o = cnfc phi2 in let o1 = cnfc phi1 in FAnd_wi (o1, o)
  | phi1 -> phi1
  end

let t (phi: formula) : formula_wi = cnfc (nnfc (impl_free phi))
\end{why3}

\newpage

\section{CPS Version}
\label{appendix:cps}

\begin{why3} 
let rec impl_free_cps (phi: formula) (k: formula_wi -> 'a ) : 'a
 = match phi with
  | FNeg phi1 -> impl_free_cps phi1 (fun con -> k (FNeg_wi con))
  | FOr phi1 phi2 -> impl_free_cps phi1 (fun con -> impl_free_cps phi2 (fun con1 -> k (FOr_wi con con1)))
  | FAnd phi1 phi2 -> impl_free_cps phi1 (fun con -> impl_free_cps phi2 (fun con1 -> k (FAnd_wi con con1)))
  | FImpl phi1 phi2 -> impl_free_cps phi1 (fun con -> impl_free_cps phi2 (fun con1 -> k (FOr_wi (FNeg_wi con) con1)))
  | FConst phi -> k (FConst_wi phi)
  | FVar phi -> k (FVar_wi phi)
end

let impl_free_main (phi: formula) : formula_wi
= impl_free_cps phi (fun x -> x)

let rec nnfc_cps (phi: formula_wi) (k: formula_wi -> 'a) : 'a
= match phi with
  | FNeg_wi (FNeg_wi phi1) -> nnfc_cps phi1 (fun con -> k con)
  | FNeg_wi (FAnd_wi phi1 phi2) -> nnfc_cps (FNeg_wi phi1) (fun con -> nnfc_cps (FNeg_wi phi2) (fun con1 -> k (FOr_wi con con1)))
  | FNeg_wi (FOr_wi phi1 phi2) -> nnfc_cps (FNeg_wi phi1) (fun con -> nnfc_cps (FNeg_wi phi2) (fun con1 -> k (FAnd_wi con con1)))
  | FOr_wi phi1 phi2 -> nnfc_cps phi1 (fun con -> nnfc_cps phi2 (fun con1 -> k (FOr_wi con con1)))
  | FAnd_wi phi1 phi2 -> nnfc_cps phi1 (fun con -> nnfc_cps phi2 (fun con1 -> k (FAnd_wi con con1)))
  | phi -> k (phi)
end

let nnfc_main (phi: formula_wi) : formula_wi
= nnfc_cps phi (fun x -> x)

let rec distr_cps (phi1 phi2: formula_wi) (k: formula_wi -> 'a) : 'a
= match phi1, phi2 with
  | FAnd_wi phi11 phi12, phi2 -> distr_cps phi11 phi2 (fun con -> distr_cps phi12 phi2 (fun con1 -> k (FAnd_wi con con1)))
  | phi1, FAnd_wi phi21 phi22 -> distr_cps phi1 phi21 (fun con -> distr_cps phi1 phi22 (fun con1 -> k (FAnd_wi con con1)))
  | phi1,phi2 -> k (FOr_wi phi1 phi2)
    end

let distr_main (phi1 phi2: formula_wi) : formula_wi
= distr_cps phi1 phi2 (fun x -> x)

let rec cnfc_cps (phi: formula_wi) (k: formula_wi -> 'a) : 'a
= match phi with
  | FOr_wi phi1 phi2 -> cnfc_cps phi1 (fun con -> cnfc_cps phi2 (fun con1 -> distr_cps con con1 k))
  | FAnd_wi phi1 phi2 -> cnfc_cps phi1 (fun con -> cnfc_cps phi2 (fun con1 -> k (FAnd_wi con con1)))
  | phi -> k (phi)
end

let cnfc_main (phi: formula_wi) : formula_wi
= cnfc_cps phi (fun x -> x)
\end{why3}

\newpage

\section{Correction of functions in CPS}
\label{appendix:provacps}

\subsection{NNFC}

\begin{why3}
let rec nnfc_cps (phi: formula_wi) (k: formula_wi -> 'a) : 'a
  variant { size phi }
  ensures { result = k (nnfc phi) }
= ...
    
let nnfc_main (phi: formula_wi) : formula_wi
  ensures { forall v. eval_wi v phi = eval_wi v result } 
  ensures { wf_negations_of_literals result }
= nnfc_cps phi (fun x -> x)
\end{why3}

\subsection{CNFC}

\begin{why3}
let rec cnfc_cps (phi: formula_wi) (k: formula_wi -> 'a) : 'a
  requires{ wf_negations_of_literals phi }
  variant { phi }
  ensures{ result = k (cnfc phi)}
= ...

let cnfc_main (phi: formula_wi) : formula_wi
  requires{ wf_negations_of_literals phi }
  ensures{ forall v. eval_wi v phi = eval_wi v result }
  ensures{ wf_negations_of_literals result }
  ensures{ wf_conjunctions_of_disjunctions result}
= cnfc_cps phi (fun x -> x)
\end{why3}

\newpage

\section{Defunctionalized Version}
\label{appendix:desfun}

\subsection{Types}
\begin{why3} 
type impl_kont =
  | KImpl_Id
  | KImpl_Neg impl_kont formula
  | KImpl_OrLeft formula impl_kont
  | KImpl_OrRight impl_kont formula_wi
  | KImpl_AndLeft formula impl_kont
  | KImpl_AndRight impl_kont formula_wi
  | KImpl_ImplLeft formula impl_kont
  | KImpl_ImplRight impl_kont formula_wi

type nnfc_kont =
  | Knnfc_id
  | Knnfc_negneg nnfc_kont formula_wi
  | Knnfc_negandleft formula_wi nnfc_kont
  | Knnfc_negandright nnfc_kont formula_wi
  | Knnfc_negorleft formula_wi nnfc_kont
  | Knnfc_negorright nnfc_kont formula_wi
  | Knnfc_andleft formula_wi nnfc_kont
  | Knnfc_andright nnfc_kont formula_wi
  | Knnfc_orleft formula_wi nnfc_kont
  | Knnfc_orright nnfc_kont formula_wi
    
type distr_kont =
  | KDistr_Id
  | KDistr_Left formula_wi formula_wi distr_kont
  | KDistr_Right distr_kont formula_wi
    
type cnfc_kont =
  | KCnfc_Id
  | KCnfc_OrLeft formula_wi cnfc_kont
  | KCnfc_OrRight cnfc_kont formula_wi
  | KCnfc_AndLeft formula_wi cnfc_kont
  | KCnfc_AndRight cnfc_kont formula_wi
\end{why3}

\subsection{Impl\_Free Function}

\begin{why3}
let rec impl_free_desf_cps (phi: formula) (k: impl_kont) : formula_wi
= match phi with
  | FNeg phi1 -> impl_free_desf_cps phi1 (KImpl_Neg k phi1)
  | FOr phi1 phi2 -> impl_free_desf_cps phi1 (KImpl_OrLeft phi2 k)
  | FAnd phi1 phi2 -> impl_free_desf_cps phi1 (KImpl_AndLeft phi2 k)
  | FImpl phi1 phi2 -> impl_free_desf_cps phi1 (KImpl_ImplLeft phi2 k)
  | FConst phi -> impl_apply (FConst_wi phi) k
  | FVar phi -> impl_apply (FVar_wi phi) k
end

with impl_apply (phi: formula_wi) (k: impl_kont) : formula_wi
= match k with
  | KImpl_Id -> phi
  | KImpl_Neg k phi1 -> impl_apply (FNeg_wi phi) k
  | KImpl_OrLeft phi1 k -> impl_free_desf_cps phi1 (KImpl_OrRight k phi)
  | KImpl_OrRight k phi2 -> impl_apply (FOr_wi phi2 phi) k
  | KImpl_AndLeft phi1 k -> impl_free_desf_cps phi1 (KImpl_AndRight k phi)
  | KImpl_AndRight k phi2 -> impl_apply (FAnd_wi phi2 phi) k
  | KImpl_ImplLeft phi1 k -> impl_free_desf_cps phi1 (KImpl_ImplRight k phi)
  | KImpl_ImplRight k phi2-> impl_apply (FOr_wi (FNeg_wi phi2) phi) k
end

let rec impl_desf_main (phi:formula) : formula_wi
= impl_free_desf_cps phi KImpl_Id

\end{why3}

\subsection{NNFC Function}

\begin{why3}
let rec nnfc_desf_cps (phi: formula_wi) (k: nnfc_kont) : formula_wi
= match phi with
  | FNeg_wi (FNeg_wi phi1) -> nnfc_desf_cps phi1 (Knnfc_negneg k phi1)
  | FNeg_wi (FAnd_wi phi1 phi2) -> nnfc_desf_cps (FNeg_wi phi1) (Knnfc_negandleft phi2 k)
  | FNeg_wi (FOr_wi phi1 phi2) -> nnfc_desf_cps (FNeg_wi phi1) (Knnfc_negorleft phi2 k)
  | FOr_wi phi1 phi2 -> nnfc_desf_cps phi1 (Knnfc_orleft phi2 k)
  | FAnd_wi phi1 phi2 -> nnfc_desf_cps phi1 (Knnfc_andleft phi2 k)
  | phi -> nnfc_apply phi k
end

with nnfc_apply (phi: formula_wi) (k: nnfc_kont) : formula_wi
= match k with
  | Knnfc_id -> phi
  | Knnfc_negneg k phi1 -> nnfc_apply phi k
  | Knnfc_negandleft phi1 k -> nnfc_desf_cps (FNeg_wi phi1) (Knnfc_negandright k phi)
  | Knnfc_negandright k phi2 -> nnfc_apply (FOr_wi phi2 phi) k
  | Knnfc_negorleft phi1 k -> nnfc_desf_cps (FNeg_wi phi1) (Knnfc_negorright k phi)
  | Knnfc_negorright k phi2 -> nnfc_apply (FAnd_wi phi2 phi) k
  | Knnfc_andleft phi1 k -> nnfc_desf_cps phi1 (Knnfc_andright k phi)
  | Knnfc_andright k phi2 -> nnfc_apply (FAnd_wi phi2 phi) k
  | Knnfc_orleft phi1 k -> nnfc_desf_cps phi1 (Knnfc_orright k phi)
  | Knnfc_orright k phi2 -> nnfc_apply (FOr_wi phi2 phi) k
end

let nnfc_desf_main (phi: formula_wi) : formula_wi
= nnfc_desf_cps phi Knnfc_id
\end{why3}

\subsection{Distr Function}

\begin{why3}
let rec distr_desf_cps (phi1 phi2: formula_wi) (k: wf_distr_kont) : formula_wi
= match phi1,phi2 with
  | FAnd_wi phi11 phi12, phi2 ->
        distr_desf_cps phi11 phi2 { distr_k = KDistr_Left phi12 phi2 k.distr_k }
  | phi1, FAnd_wi phi21 phi22 ->
        distr_desf_cps phi1 phi21 { distr_k = KDistr_Left phi1 phi22 k.distr_k }
  | phi1,phi2 -> distr_apply (FOr_wi phi1 phi2) k
end

with distr_apply (phi: formula_wi) (k: wf_distr_kont) : formula_wi
= match k.distr_k with
  | KDistr_Id -> phi
  | KDistr_Left phi1 phi2 k ->
        distr_desf_cps phi1 phi2 { distr_k = KDistr_Right k phi }
  | KDistr_Right k phi1 ->
        distr_apply (FAnd_wi phi1 phi) { distr_k = k }
end

let distr_desf_main (phi1 phi2: formula_wi) : formula_wi
= distr_desf_cps phi1 phi2 { distr_k = KDistr_Id }
\end{why3}

\subsection{CNFC Function}

\begin{why3}
let rec cnfc_desf_cps (phi: formula_wi) (k: wf_cnfc_kont) : formula_wi
= match phi with
  | FOr_wi phi1 phi2 ->
        cnfc_desf_cps phi1 { cnfc_k = KCnfc_OrLeft phi2 k.cnfc_k }
  | FAnd_wi phi1 phi2 ->
        cnfc_desf_cps phi1 { cnfc_k = KCnfc_AndLeft phi2 k.cnfc_k }
  | phi -> cnfc_apply phi k
end

with cnfc_apply (phi: formula_wi) (k: wf_cnfc_kont) : formula_wi
= match k.cnfc_k with
  | KCnfc_Id -> phi
  | KCnfc_OrLeft phi1 k ->
        cnfc_desf_cps phi1 { cnfc_k = KCnfc_OrRight k phi }
  | KCnfc_OrRight k phi2 ->
        cnfc_apply (distr_desf_cps phi2 phi { distr_k = KDistr_Id })
          { cnfc_k = k }
  | KCnfc_AndLeft phi1 k ->
        cnfc_desf_cps phi1 { cnfc_k = KCnfc_AndRight k phi }
  | KCnfc_AndRight k phi2 ->
        cnfc_apply (FAnd_wi phi2 phi) { cnfc_k = k }
end

let cnfc_desf_main (phi: formula_wi) : formula_wi
= cnfc_desf_cps phi { cnfc_k = KCnfc_Id }
\end{why3}

\newpage

\section{Correction of defunctionalized functions}
\label{appendix:provadesfun}

\subsection{NNFC}

\begin{why3}
predicate nnfc_post (k: nnfc_kont) (phi result: formula_wi)
= match k with
  | Knnfc_id -> let x = phi in x = result
  | Knnfc_negneg k phi1 -> let neg = phi in nnfc_post k phi result
  | Knnfc_negandleft phi1 k -> let hl = phi in nnfc_post k (FOr_wi phi (nnfc (FNeg_wi phi1))) result
  | Knnfc_negandright k phi2 -> let hr = phi in nnfc_post k (FOr_wi phi2 hr) result
  | Knnfc_negorleft phi1 k -> let hl = phi in nnfc_post k (FAnd_wi phi (nnfc (FNeg_wi phi1))) result
  | Knnfc_negorright k phi2 ->let hr = phi in nnfc_post k (FAnd_wi phi2 hr) result
  | Knnfc_andleft phi1 k -> let hl = phi in nnfc_post k (FAnd_wi phi (nnfc phi1)) result
  | Knnfc_andright k phi2 -> let hr = phi in nnfc_post k (FAnd_wi phi2 hr) result
  | Knnfc_orleft phi1 k -> let hl = phi in nnfc_post k (FOr_wi phi (nnfc phi1)) result
  | Knnfc_orright k phi2 -> let hr = phi in nnfc_post k (FOr_wi phi2 hr) result
end
  
let rec nnfc_desf_cps (phi: formula_wi) (k: nnfc_kont) : formula_wi
  ensures{ nnfc_post k (nnfc phi) result }
= ...

with nnfc_apply (phi: formula_wi) (k: nnfc_kont) : formula_wi
  ensures{ nnfc_post k phi result }
= ...
\end{why3}

\subsection{CFNC}

\begin{why3}
predicate cnfc_post (k: cnfc_kont) (phi result: formula_wi)
= match k with
  | KCnfc_Id -> let x = phi in x = result
  | KCnfc_OrLeft phi1 k -> let hl = phi in cnfc_post k (distr hl (cnfc phi1)) result
  | KCnfc_OrRight k phi2 -> let hr = phi in cnfc_post k (distr phi2 hr) result
  | KCnfc_AndLeft phi1 k -> let hl = phi in cnfc_post k (FAnd_wi phi (cnfc phi1)) result
  | KCnfc_AndRight k phi2 -> let hr = phi in cnfc_post k (FAnd_wi phi2 hr) result
end
  
let rec cnfc_desf_cps (phi: formula_wi) (k: wf_cnfc_kont) : formula_wi
  requires{ wf_negations_of_literals phi }
  ensures{ cnfc_post k.cnfc_k (cnfc phi) result }
= ...

with cnfc_apply (phi: formula_wi) (k: wf_cnfc_kont) : formula_wi
  requires{ wf_negations_of_literals phi }
  requires { wf_conjunctions_of_disjunctions phi }
  ensures{ cnfc_post k.cnfc_k phi result }
= ...
\end{why3}

\subsection{Distr Defunctionalized Function Specification}
\begin{why3}
predicate wf_distr_kont (phi: distr_kont) 
= match phi with
  | KDistr_Id -> true
  | KDistr_Left phi1 phi2 k ->
        wf_negations_of_literals phi1 /\ wf_conjunctions_of_disjunctions phi1 /\
        wf_negations_of_literals phi2 /\ wf_conjunctions_of_disjunctions phi2 /\
        wf_distr_kont k
  | KDistr_Right k phi ->
        wf_negations_of_literals phi /\ wf_conjunctions_of_disjunctions phi /\ wf_distr_kont k
end

type wf_distr_kont = {
    distr_k: distr_kont;
} invariant { wf_distr_kont distr_k }
by { distr_k = KDistr_Id }
\end{why3}\textbf{}

%% file: main.bbl
\begin{thebibliography}{10}
\providecommand{\url}[1]{\texttt{#1}}
\providecommand{\urlprefix}{URL }
\providecommand{\doi}[1]{https://doi.org/#1}

\bibitem{gitlab}
{Add injectivity for type invariant ({\#}287) · Why3 Issues},
  \url{https://gitlab.inria.fr/why3/why3/issues/287}

\bibitem{factor}
{FACTOR: Functional ApproaCh Teaching pOrtuguese couRses},
  \url{http://ctp.di.fct.unl.pt/FACTOR/}

\bibitem{danvy2003functionalAM}
Ager, M.S., Biernacki, D., Danvy, O., Midtgaard, J.: A functional
  correspondence between evaluators and abstract machines. In: Proceedings of
  the International Conference on Principles and Practice of Declarative
  Programming (2003)

\bibitem{mathlogicbenari}
Ben{-}Ari, M.: {Mathematical Logic for Computer Science, 3rd Edition}. Springer
  (2012), \url{https://doi.org/10.1007/978-1-4471-4129-7}

\bibitem{inttologicenderton}
Enderton, H.B.: A mathematical introduction to logic. Academic Press (1972)

\bibitem{filliatre2013why3}
Filli{\^{a}}tre, J., Paskevich, A.: {Why3 - Where Programs Meet Provers}. In:
  Programming Languages and Systems. Lecture Notes in Computer Science,
  Springer, \url{https://doi.org/10.1007/978-3-642-37036-6\_8}

\bibitem{logichamilton}
Hamilton, A.G.: {Logic for mathematicians}. Cambridge University Press (1988)

\bibitem{hornalgorithm}
Horn, A.: {On Sentences Which are True of Direct Unions of Algebras}, vol.~16
  (1951), \url{https://doi.org/10.2307/2268661}

\bibitem{huth2004logic}
Huth, M., Ryan, M.D.: Logic in computer science - modelling and reasoning about
  systems {(2.} ed.). Cambridge University Press (2004)

\bibitem{leroy2014ocaml}
Leroy, X., Doligez, D., Frisch, A., Garrigue, J., R{\'e}my, D., Vouillon, J.:
  {The OCaml system release 4.07: Documentation and user's manual}. Intern
  report, {Inria} (2018), \url{https://hal.inria.fr/hal-00930213}

\bibitem{mathlogicmendelson}
Mendelson, E.: Introduction to mathematical logic {(3.} ed.). Chapman and Hall
  (1987)

\bibitem{pereira2019desfuncionalizar}
Pereira, M.: {Desfuncionalizar para Provar}. CoRR  \textbf{abs/1905.08368}
  (2019), \url{http://arxiv.org/abs/1905.08368}

\bibitem{reynolds1972definitional}
Reynolds, J.C.: {Definitional Interpreters for Higher-Order Programming
  Languages}. vol.~11, pp. 363--397 (1998),
  \url{https://doi.org/10.1023/A:1010027404223}

\bibitem{sabry1993reasoning}
Sabry, A., Felleisen, M.: {Reasoning about Programs in Continuation-Passing
  Style}. Lisp and Symbolic Computation  \textbf{6}(3-4),  289--360 (1993)

\end{thebibliography}
